\documentclass[%
 aip,
 amsmath,amssymb,
 reprint,%
]{revtex4-1}

\usepackage{graphicx}% Include figure files
\usepackage{dcolumn}% Align table columns on decimal point
\usepackage{bm}% bold math
\usepackage[utf8]{inputenc}
\usepackage[T1]{fontenc}
\usepackage{mathptmx}
\usepackage{etoolbox}

\makeatletter
\def\@email#1#2{%
 \endgroup
 \patchcmd{\titleblock@produce}
  {\frontmatter@RRAPformat}
  {\frontmatter@RRAPformat{\produce@RRAP{*#1\href{mailto:#2}{#2}}}\frontmatter@RRAPformat}
  {}{}
}%
\makeatother
\begin{document}

\preprint{AIP/123-QED}

\title[Bounce-averaged drifts]{Bounce-averaged drifts:\\ Equivalent definitions, numerical implementations, and example cases}
% Force line breaks with \\
\author{R.J.J. Mackenbach}
\affiliation{Eindhoven University of Technology, 5612 AZ, Eindhoven, Netherlands}
\affiliation{Max Planck Institute for Plasma Physics, 17491 Greifswald, Germany}
\email{r.j.j.mackenbach@tue.nl}
\author{J.M. Duff}%
\affiliation{University of Wisconsin-Madison, Madison Wisconsin 53706, USA}
\author{M.J. Gerard}
\affiliation{University of Wisconsin-Madison, Madison Wisconsin 53706, USA}
\author{J.H.E. Proll}
\affiliation{Eindhoven University of Technology, 5612 AZ, Eindhoven, Netherlands}
\affiliation{Max Planck Institute for Plasma Physics, 17491 Greifswald, Germany}
\author{P. Helander}
\affiliation{Max Planck Institute for Plasma Physics, 17491 Greifswald, Germany}
\author{C.~C.~Hegna}
\affiliation{University of Wisconsin-Madison, Madison Wisconsin 53706, USA}

\date{\today}% It is always \today, today,

\begin{abstract}
In this article we provide various analytical and numerical methods for calculating the average drift of magnetically trapped particles across field lines in complex geometries, and we compare these methods against each other. To evaluate bounce-integrals, we introduce a generalisation of the trapezoidal rule which is able to circumvent integrable singularities. We contrast this method with more standard quadrature methods in a parabolic magnetic well and find that the computational cost is significantly lower for the trapezoidal method, though at the cost of accuracy. With numerical routines in place, we next investigate conditions on particles which cross the computational boundary, and we find that important differences arise for particles affected by this boundary, which can depend on the specific implementation of the calculation. Finally, we investigate the bounce-averaged drifts in the optimized stellarator NCSX. From investigating the drifts, one can readily deduce important properties, such as what subset of particles can drive trapped-particle modes, and in what regions radial drifts are most deleterious to the stability of such modes. 
\end{abstract}

\maketitle

\section{Introduction}
The total drift that a particle trapped in a magnetic well experiences between two consecutive bounce points, henceforth referred to as the bounce-averaged drift (BAD), plays an important role in plasma physics. In stellarator physics, for example, an enormous amount of work has been devoted to the construction of equilibria in which the BAD in the radial direction is close to zero, as is the case in quasi-axisymmetric, quasi-helically symmetric, and quasi-isodynamic fields \cite{boozer1983transport,rodriguez2020necessary,beidler2021demonstration,landreman2022magnetic,goodman2022constructing}. The BAD in the direction of the Clebsch angle (often referred to as the binormal drift) on the other hand, plays a central role in determining linear stability of the gyrokinetic trapped electron mode \cite{proll2012resilience,helander2013collisionless,helander2017available}. As such, it is of interest to have accurate methods of calculating these quantities in generic equilibria. \par 
Though many authors have implemented methods to calculate such quantities \cite{white1984hamiltonian,roach1995trapped,marinoni2009effect,proll2015tem,gan2021self,stephens2021quasilinear,mackenbach2022available}, these are not always valid for general equilibria, or they may refer to codes which have been superseded by newer codes. There are furthermore equivalent methods of calculating these drifts, some of which may be easier to implement, depending on the quantities given \cite{helander2005collisional,hegna2015effect}. Accordingly, the goal of the current paper is to provide an overview of various analytical and numerical methods of calculating BADs, in order to facilitate the choice of method suited for any particular problem at hand. Using this framework, we furthermore investigate the BADs in the NCSX device, and we calculate the effect of a radial electric field on such drifts.

\section{Methods of calculating the BAD}
In this section, we highlight several analytical methods by which the BAD may be calculated using well-known results from the literature \cite{helander2014theory}. We start by considering the local instantaneous drift velocity of a guiding centre across the magnetic field,
\begin{equation}
    \mathbf{v}_D = \frac{\mathbf{E} \times \mathbf{B}}{B^2} + \frac{ v_\perp^2}{2 \Omega}\frac{\mathbf{B} \times \nabla B}{B^2} + \frac{v_\parallel^2}{\Omega} \frac{\mathbf{B} \times \bm{\kappa}}{B},
\end{equation}
where $\Omega = q B / m$ is the gyration frequency of a particle with mass $m$ and charge $q$, and we have neglected polarisation drifts \cite{blank2004guiding}. Furthermore $v_\parallel$ and $v_\perp$ are the parallel and perpendicular velocities of the particle with respect to the magnetic field. It is often convenient to express this drift in terms of a pitch-angle variable $\lambda \equiv \mu B_0/H$, where $H = m (v_\|^2 + v_\perp^2)/2 + \Phi$ is the total (kinetic + potential) energy and $B_0$ is some reference magnetic field strength. In terms of this quantity, the drift can be expressed as
\begin{equation}
\begin{aligned}
    \mathbf{v}_D = &\frac{\mathbf{E} \times \mathbf{B}}{B^2} + \\
    &\frac{H}{q B_0} \left[  \lambda \frac{\mathbf{B} \times \nabla B}{B^2} + 2 \left(\hat{B}^{-1}-\lambda - \frac{\Phi}{H \hat{B}} \right) \frac{\mathbf{B} \times \bm{\kappa}}{B} \right],
\end{aligned}
\end{equation}
where $\hat{B} = B/B_0$. It is of interest to note that, when MHD equilibrium conditions are satisfied, there is a simple relation between the curvature and gradient drifts. By invoking MHD force balance on the definition of the curvature vector $\bm{\kappa} = - \mathbf{b} \times ( \nabla \times \mathbf{b} )$ we find
\begin{equation}
    \bm{\kappa} =  \frac{\mu_0 \nabla p}{B^2} + \frac{\nabla_\perp B}{B}.
\end{equation}
If we substitute the above relationship into the expression for the curvature drift, the relation becomes
\begin{equation}
    \frac{\mathbf{B} \times \bm{\kappa}}{B} = \mathbf{b} \times \frac{\mu_0 \nabla p}{B^2} + \frac{\mathbf{B} \times \nabla B}{B^2},
    \label{eq:relation-gradandcurv}
\end{equation}
with $\mathbf{b}=\mathbf{B}/B$. We now shift our attention to the parallel motion, which obeys
\begin{equation}
    m \frac{\mathrm{d} v_\parallel }{\mathrm{d} t} = - \nabla_\parallel ( \Phi + \mu B ).
\end{equation}
Here, $\Phi$ is the electric potential so that $\mathbf{F} = - \nabla \Phi = q \bf E$ denotes the electric force, $\nabla_\parallel = {\bf b} \cdot \nabla $ is the gradient parallel to the magnetic field, and $\mu = m v_\perp^2 /(2B)$ is the magnetic moment, the first adiabatic invariant. It should be noted that these expressions for the drifts are only valid if the magnetic field varies slowly on the length scale of the gyroradius and the electric field is not too large.
\subsection{Particle orbit tracing} \label{sec:orbit}
We shall first look into the most straightforward method of calculating the BADs, namely particle orbit tracing. These methods solve the gyro-averaged equations of motion,
\begin{equation}
    \frac{\mathrm{d} \mathbf{r}}{\mathrm{d} t} =\mathbf{v}_D + v_\parallel \mathbf{b}.
\end{equation}
A wide variety of methods are available to calculate particle orbits accurately, including algorithms which are Hamiltonian and conserve symplectic invariants \cite{albert2020accelerated}. Such methods are particularly valuable if one wishes to compute orbits over long times (e.g. many bounces), during which orbits may transition between different trapping classes, which in turn can be important for neoclassical transport \cite{paul2022energetic}. In order to find the BAD of the above expression, one simply calculates the bounce time, i.e. the total duration of a bounce-motion between two consecutive bounce points (i.e. where $v_\parallel = 0$), and divides the total experienced drift by this quantity. Denoting the bounce-averaging operator by means of angular brackets, we thus have that this averaged drift becomes
\begin{equation}
    \langle \mathbf{v}_D \rangle \equiv \frac{ \int \mathbf{v}_D \; \mathrm{d} t }{\int \mathrm{d} t}.
\end{equation}
The above quantities can readily be extracted from a particle orbit tracing codes, and since such codes have been benchmarked extensively they will not be discussed further in the current paper. These methods are likely the most computationally expensive of all the methods that will be presented. 
\subsection{Direct averaging} \label{sec:direct-averaging}
One can also calculate the BAD straight from its definition. This can be useful for gyrokinetic codes, especially local ones in which the projections of the drift are often given as functions of some field line following coordinate, which we will denote as $\zeta$. This field line following coordinate could, for example, be a linear combination of toroidal and poloidal straight-field line angles, typically denoted as $\varphi$ and $\theta$ respectively. With such a field-line following coordinate, the parallel derivative becomes
\begin{equation}
    \mathbf{B} \cdot \nabla = \left( \mathbf{B} \cdot \nabla \zeta \right) \frac{\partial}{\partial \zeta}.
\end{equation}
\par 
Suppose one is given the following projection of the drift
\begin{equation}
    \mathbf{v}_D \cdot \nabla x,
\end{equation}
where $x$ is a placeholder for either a radial coordinate such as toroidal flux $\psi$, or a binormal coordinate such a the Clebsch angle $\alpha = \theta - \iota \varphi$, where $\iota$ denotes the rotational transform. One can then compute the bounce-average from its definition
\begin{equation}
    \langle \mathbf{v}_D \cdot \nabla x \rangle \equiv \frac{ \int \mathbf{v}_D \cdot \nabla x \; \mathrm{d} t }{\int \mathrm{d} t}.
\end{equation}
Since to lowest order the parallel dynamics dominate, we may approximate the infinitesimal $\mathrm{d} t$ as
\begin{equation}
    \mathrm{d} t = \frac{\mathrm{d} \ell}{v_\parallel},
\end{equation}
where $\ell$ is the arc length along a magnetic field line. The parallel velocity is related to the energy by\begin{equation}
\begin{aligned}
    & H = \frac{1}{2}mv_\parallel^2 + \mu B + \Phi \implies \\
    & v_\parallel = \pm \sqrt{\frac{2H}{m}} \sqrt{1 - \lambda \hat{B} - \frac{\Phi}{H}}.
\end{aligned}
\end{equation}
Note that the bounce-averaging operator is invariant under $v_\parallel \mapsto C \cdot v_\parallel$ for any constant $C$, and hence the BAD becomes
\begin{equation}
    \langle \mathbf{v}_D \cdot \nabla x \rangle = \frac{\int \mathrm{d} \ell \; (\mathbf{v}_D \cdot \nabla x) \left(1 - \lambda \hat{B} - \frac{\Phi}{H} \right)^{-1/2}  }{\int \mathrm{d} \ell  \left(1 - \lambda \hat{B} - \frac{\Phi}{H} \right)^{-1/2} }.
    \label{eq:direct-averaging-integral}
\end{equation}
Finally one should take care in expressing these quantities in terms of the field line following coordinate $\zeta$. The relation between arc-length and this field line following coordinate can readily be found by investigating the projection of the magnetic field along $\zeta$,
\begin{equation}
    \mathbf{B} \cdot \nabla \zeta = \left( \nabla \psi \times \nabla \alpha \right) \cdot \nabla \zeta,
\end{equation}
where we have made use of the Clebsch representation of the magnetic field $\mathbf{B} = \nabla \psi \times\nabla \alpha$.
Since $\zeta$ is a field line following coordinate, we set $\zeta = \zeta(\ell)$ the above equation becomes
\begin{equation}
    \frac{\mathrm{d} \zeta}{\mathrm{d} \ell} = \frac{\left( \nabla \psi \times \nabla \alpha \right) \cdot \nabla \zeta}{B}.
\end{equation}
As an example case, if one chooses $\nabla \zeta = a_\theta \nabla \theta + a_\varphi \nabla \varphi$ with $a_\theta$ and $a_\varphi$ being constants which satisfy $a_\theta \iota + a_\varphi \neq 0 $ (i.e. the field-line following coordinate should not be perpendicular to $\mathbf{B}$), the equation becomes
\begin{equation}
\begin{aligned}
    \frac{\mathrm{d} \zeta}{\mathrm{d} \ell} =& (a_\theta \iota + a_\varphi) \frac{\left( \nabla \psi \times \nabla \theta \right) \cdot \nabla \varphi}{B}.
\end{aligned}
\end{equation}
This expression is particularly simple in Boozer coordinates \cite{boozer1981plasma,helander2014theory}, as the inverse Jacobian $(\nabla \psi \times \nabla \theta ) \cdot \nabla \varphi$ becomes proportional to $B^2$. This simplifies the BAD greatly, because the bounce averaging operator is invariant under $\mathrm{d}\zeta/\mathrm{d} \ell \mapsto C \cdot \mathrm{d}\zeta/\mathrm{d} \ell $. In Boozer coordinates, one can then replace
\begin{equation}
    \frac{\mathrm{d} \ell}{\mathrm{d} \zeta} \mapsto \frac{1}{B}
\end{equation}
in equation \eqref{eq:direct-averaging-integral}, as long as $\zeta$ is a linear combination of Boozer angles. We thus see that field-lines spend more of their arc-length in regions of low magnetic field strength in the Boozer plane. \par
Returning to the general case, one can now express the BAD in terms of an arbitrary field line following coordinate, and we thus have
\begin{equation}
    \langle \mathbf{v}_D \cdot \nabla x  \rangle = \frac{\int \mathrm{d} \zeta \; (\mathbf{v}_D \cdot \nabla x) \frac{\mathrm{d} \ell}{\mathrm{d} \zeta} \left(1 - \lambda \hat{B} - \frac{\Phi}{H} \right)^{-1/2}  }{\int \mathrm{d} \zeta  \frac{\mathrm{d} \ell}{\mathrm{d} \zeta} \left(1 - \lambda \hat{B} - \frac{\Phi}{H} \right)^{-1/2} }.
\end{equation}
This expression in now in a form accessible for many codes such as GIST \cite{xanthopoulos2009geometry}, GENE \cite{jenko2001critical}, or simsopt \cite{landreman2021simsopt}. One should keep in mind that these expressions are only accurate in the limit of small $\rho_*=\rho_g/a$, where $\rho_g$ is the gyroradius and $a$ is the minor radius.

\subsection{Geometric approach} \label{sec:geom-approach}
One can also take a more geometric approach in calculating the BADs, namely by investigating its relation to the second adiabatic invariant, denoted as $\mathcal{J}$, which is given by
\begin{equation}
    \mathcal{J} = \int m v_\parallel \mathrm{d} \ell.
\end{equation}
From the gyro-averaged Lagrangian one can find the following relations \cite{helander2014theory}
\begin{subequations}
\label{eq:whole}
\begin{equation}
\left( \frac{\partial \mathcal{J}}{\partial \psi} \right)_{H,\mu,\alpha} = - q \Delta \alpha,\label{subeq_dJ:1}
\end{equation}
\begin{eqnarray}
\left( \frac{\partial \mathcal{J}}{\partial \alpha} \right)_{H,\mu,\psi} = + q \Delta \psi \label{subeq_dJ:2},
\end{eqnarray}
\begin{eqnarray}
\left( \frac{\partial \mathcal{J}}{\partial H} \right)_{\mu,\psi,\alpha} = \tau_b \label{subeq_dJ:3},
\end{eqnarray}
\end{subequations}
where $\Delta \alpha$ and $\Delta \psi$ are the total excursion in $\alpha$ and $\psi$ after a full bounce-motion, and $\tau_b$ is the total bounce-time of that bounce motion. Given these expressions, the BAD in $\alpha$ and $\psi$ is readily found to be
\begin{subequations}
\label{eq:whole}
\begin{equation}
\langle \mathbf{v}_D \cdot \nabla \alpha \rangle = - \frac{1}{q} \left( \frac{\partial \mathcal{J}}{\partial \psi} \right)_{H,\mu,\alpha} \Bigg/ \left( \frac{\partial \mathcal{J}}{\partial H} \right)_{\mu,\psi,\alpha},\label{subeq:1}
\end{equation}
\begin{eqnarray}
\langle \mathbf{v}_D \cdot \nabla \psi \rangle = + \frac{1}{q} \left( \frac{\partial \mathcal{J}}{\partial \alpha} \right)_{H,\mu,\psi} \Bigg/ \left( \frac{\partial \mathcal{J}}{\partial H} \right)_{\mu,\psi,\alpha} \label{subeq:2}.
\end{eqnarray}
\end{subequations}
The main difficulty with the above expressions is that one needs to take care in expressing these derivatives in their chosen coordinate system. For example, the equations are exceptionally simple if one expresses the various functions in terms of the Clebsch coordinates $(\psi,\alpha,\ell)$. \par
Following arguments based on the the local 3D equilibrium model, \cite{hegna2000local,hegna2015effect,Gerard_2023} equations \eqref{subeq_dJ:1} and \eqref{subeq_dJ:3} can similarly be expressed in straight-field line angular coordinates. Doing so, results in the following equations for the derivatives
% \begin{subequations}
% \label{eq:whole}
% \begin{equation}
% \begin{split}
% \left( \frac{\partial \mathcal{J}}{\partial \psi} \right)_{H,\mu,\alpha} =& \frac{H}{N - \iota M} \int \frac{\sqrt{g}B}{v_{\parallel}} \times \\
% & \left[ \frac{2\hat{v}_{\parallel}^2M}{N - \iota M} \frac{d\iota}{d\psi} \right.
% - \\ 
% & \left( 1 + \hat{v}_{\parallel}^2 \right) \left( \frac{\kappa_{\mathrm{n}}}{|\nabla \psi|} + D\kappa_{\mathrm{g}} \frac{|\nabla \psi|}{B} \right) + \\
% & \left. \left( 1 - \hat{v}_{\parallel}^2 \right) \frac{\mu_0}{B^2} \frac{dp}{d\psi}\right] d\eta, \label{local3D_dJ:1}
% \end{split}
% \end{equation}
% \begin{eqnarray}
% \left( \frac{\partial \mathcal{J}}{\partial H} \right)_{\mu,\psi,\alpha} = \frac{1}{N - \iota M} \int \frac{\sqrt{g}B}{v_{\parallel}} d\eta \label{local3D_dJ:2}.
% \end{eqnarray}
% \end{subequations}
% [IN NOTATION OF MANUSCRIPT, WILL DELETE THE ABOVE EQUATION WHEN COMPARISON IS DONE]
\begin{subequations}
\label{eq:whole}
\begin{equation}
\begin{split}
& \left( \frac{\partial \mathcal{J}}{\partial \psi} \right)_{H,\mu,\alpha} \approx \sqrt{\frac{m H}{2}} \int \frac{B}{ \mathbf{B} \cdot \nabla \zeta } \frac{1}{ \sqrt{1 - \lambda \hat{B}} } \times \\
& \Bigg\{ \frac{2M( 1 - \lambda \hat{B} )}{N - \iota M} \left[ \frac{\mathrm{d}\iota}{\mathrm{d}\psi} - \frac{1}{2} \left( \frac{\kappa_{\mathrm{n}}}{|\nabla \psi|} + D\kappa_{\mathrm{g}} \frac{|\nabla \psi|}{B} \right) \right] 
- \\ 
& \left[ \frac{\kappa_{\mathrm{n}}}{|\nabla \psi|} + D\kappa_{\mathrm{g}} \frac{|\nabla \psi|}{B} \right] +  \frac{\lambda}{\hat{B}} \frac{\mu_0}{B_0^2} \frac{ \mathrm{d} p}{ \mathrm{d} \psi}\Bigg\} \mathrm{d}\zeta, \label{local3D_dJ:1}
\end{split}
\end{equation}
\begin{eqnarray}
\left( \frac{\partial \mathcal{J}}{\partial H} \right)_{\mu,\psi,\alpha} = \sqrt{\frac{m}{2H}} \int \frac{B}{ \mathbf{B} \cdot \nabla \zeta } \frac{\mathrm{d}\zeta}{\sqrt{1 - \lambda \hat{B}}}  \label{local3D_dJ:2}.
\end{eqnarray}
\end{subequations}
For the above equations we choose the field-line following coordinate to be $\zeta = N\varphi - M\theta$, where $\zeta$ is a helical angle with $M$ and $N$ the number of poloidal and toroidal field periods, respectively. For example, a quasi-axisymmetric stellarator has $N=0$ and $M=1$. Furthermore, the geodesic curvatures are defined as $\kappa_{\mathrm{n}} = \mathbf{n} \cdot \bm{\kappa}$ and $\kappa_{\mathrm{g}} = \left(\mathbf{n} \times \mathbf{b} \right) \cdot \bm{\kappa}$ respectively, with $\mathbf{n} = \nabla \psi /|\nabla \psi|$. Lastly, $D$ is related to the local magnetic shear via
\begin{equation}
s_\mathrm{loc} = \frac{|\nabla \psi|^2}{B^2} \left(\bm{B}\cdot\nabla\right) \left( \varphi\frac{d\iota}{d\psi} + D \right), \label{local3D_D_and_sloc}
\end{equation}
where $s_\mathrm{loc} = (\mathbf{b} \times \mathbf{n}) \cdot \nabla \times (\mathbf{b} \times \mathbf{n}) $ is the local magnetic shear. Importantly, equation \eqref{local3D_dJ:1} neglects terms that are a factor $(\iota\varepsilon)^2$ smaller than the terms included in the expressions, with $\varepsilon$ the inverse aspect ratio. The benefit of these expressions is that they are easy to interpret: one can readily see contributions of local shear, the curvature components, and the diamagnetic effect to the total BAD in the $\nabla\alpha$ direction.
\section{Numerical implementations}
In this section, we write down various numerical methods which may be of use in calculating the BADs. We do not focus on the numerical methods used in calculating particle orbits, as such methods have been thoroughly documented. Instead, we focus on the methods which involve solving bounce-averaging integrals. The main difficulty here is that the integrals are of the form,
\begin{equation}
    \int \frac{h(x)}{\sqrt{f(x)}} \mathrm{d} x ,
    \label{eq:b-ave-integrals-general}
\end{equation}
where the region of integration is set by the region where $f(x)>0$. The integrand becomes singular when $f(x) \rightarrow 0$, and to resolve this singularity accurately one needs to take care in choosing the numerical method. \par  
From equation \eqref{eq:direct-averaging-integral} we see that this singularity physically corresponds to the parallel velocity $v_\parallel \rightarrow 0$ at the bounce points, and thus a trapped particle can spend a significant fraction of its orbit time near these bounce points. Since the BAD is a time-average of the drift, the bounce points can account for a significant fraction of the total BAD. This effect is exacerbated near local maxima of magnetic wells, where the fraction of time a particle spends near the local maximum formally approaches unity, i.e. the particle is spending all of its orbit time at the local maximum. In such cases the BAD is fully determined by the drift at this local maximum. Accurately resolving the singular behaviour is thus of physical importance as well: if not accounted for correctly one can neglect a significant fraction of the total BAD.
\subsection{Generalisation of the trapezoidal rule} \label{eq:sec-gtrapz}
A popular choice for numerical integration is a trapezoidal method, since such a method is cheap to compute. However, near the singularity, errors can become arbitrarily large. Take, for example, the integral
\begin{equation}
    \int_\delta^{\delta+\Delta x} \frac{\mathrm{d} x}{\sqrt{x}} = 2(\sqrt{\Delta x + \delta} - \sqrt{\delta}).
    \label{eq:example-integral}
\end{equation}
A trapezoidal rule with spacing $\Delta x$ would approximate this integral as
\begin{equation}
    \int_\delta^{\delta+\Delta x} \frac{\mathrm{d} x}{\sqrt{x}} \approx \frac{\Delta x}{2} \left( \frac{1}{\sqrt{\Delta x + \delta}} + \frac{1}{\sqrt{\delta}}.  \right)
\end{equation}
If one approaches the singularity by letting $\delta \rightarrow 0^+$, the trapezoidal estimate can become arbitrarily large, whereas the true integral converges simply to $2 \sqrt{\Delta x}$. Hence, a naive trapezoidal method is undesirable. \par 
The main assumption that is broken when applying the trapezoidal method to the above integral, is that the integrand must vary only slightly from sampled location to location. This criterion is evidently broken near the singularity, resulting in the poor performance of the trapezoidal rule. It is however possible to find an analogous method by only slightly adjusting main assumptions that feed into the trapezoidal rule. Let us consider the generalized problem of estimating the integral of
\begin{equation}
    \int g(f(x)) \mathrm{d} x,
\end{equation}
where we now assume that $f(x)$ varies only slightly from location to location. More precisely, given two closely spaced samples $f(x_i)$ and $f(x_j)$ with $x_j>x_i$, we assume that the function $f(x \in [x_i,x_j])$ is well approximated by
\begin{equation}
    f(x) = f_i + \frac{x - x_i}{x_j - x_i} \left( f_j - f_i \right),
\end{equation}
where we have introduced the shorthand $f(x_k) \equiv f_k$. One could choose any integrable function $g(x)$, and analytical forms for the integral may be found if the antiderivative of $g(x)$ is known. Let us define $G'(x) = g(x)$, one can then readily show that
\begin{equation}
    \int_{x_i}^{x_j} g(f(x)) \mathrm{d}x = (x_j - x_i)\frac{G(f_j) - G(f_i)}{f_j - f_i}.
    \label{eq:gtrapz}
\end{equation}
The above equation is especially useful in situations where an integrable singularity exists due to $g(x)$. This singularity is then removed in $G(x)$, and the generalized trapezoidal rule does not suffer from the problems encountered by naive trapezoidal integration. One can furthermore set $g(x)=x$ and the normal trapezoidal rule is retrieved. For our case we will set $g(x) = 1/\sqrt{x} \implies G(x) = 2\sqrt{x}$, and the integral becomes
\begin{equation}
    \int_{x_i}^{x_j} \frac{\mathrm{d} x}{\sqrt{f(x)}} \approx 2 \frac{x_j - x_i}{\sqrt{f_i} + \sqrt{f_j}}.
\end{equation}
If one applies this generalized trapezoidal rule to the example of equation \eqref{eq:example-integral}, one retrieves the exact result. This is perhaps unsurprising, as $f(x) = x$ is exactly a linear function. One can generalize this problem further still by including the function $h(x)$ of  equation \eqref{eq:b-ave-integrals-general}, though one does need to assume that the antiderivative of $G(x)=\mathcal{G}'(x)$ is also known. We again assume that the function $h(x)$ is well approximated by
\begin{equation}
     h(x) = h_i + \frac{x - x_i}{x_j - x_i} \left( h_j - h_i \right).
\end{equation}
We can then use integration by parts to express the general integral as
\begin{equation}
\begin{aligned}
    I =& \int_{x_i}^{x_j} h(x) g(f(x))\mathrm{d} x  \\
    =& \left[ \frac{h(x)}{f'(x)} G(f(x))  \right]_{x_i}^{x_j} - \int_{x_i}^{x_j} \frac{h'(x)}{f'(x)} G(f(x)) \mathrm{d}x  \\
    =& (x_j - x_i) \frac{h_jG(f_j) - h_iG(f_i)}{f_j - f_i} - \\
    &  \frac{(x_j - x_i) (h_j - h_i)}{(f_j - f_i)^2} \left( \mathcal{G}(f_j) - \mathcal{G}(f_i) \right)
\end{aligned}
\label{eq:trapz-general}
\end{equation}
Note that equation \eqref{eq:trapz-general} is the same as equation \eqref{eq:gtrapz} if $h=1$. One can now readily estimate bounce-averaging integrals of equation \eqref{eq:b-ave-integrals-general} by setting
\begin{equation}
    g(x) = \frac{1}{\sqrt{x}}, \quad G(x) = 2\sqrt{x}, \quad \mathcal{G}(x) = \frac{4}{3} x^{3/2},
 \quad \end{equation}
and note that the singular behaviour in $I$ is removed. This trapezoidal method has one extra benefit, which is of relevance for the bounce-averaging integrals. To do the bounce-average integral one needs to numerically find the region where $f(x)>0$, which involves solving $f(x)=0$ and such a root-finding problem can be expensive. If one assumes $f(x)$ to be a piece-wise linear function between the various sampled nodes, however, the root finding becomes trivial and a computationally cheap calculation. Assuming that $f(x)$ is piece-wise linear is in line with assumptions feeding into the presented trapezoidal rule, and hence the linear root-finding and integration methods are consistent. \par 
As a final remark, we note that the presented method can be useful even in cases where $g(x)$ is not known explicitly, but the singularity type is. For example, suppose we wish to integrate the function $k(x)$ which has some $\ln(x)$ type singularity. One can then simply substitute $g(x) = \ln(x)$ and $f(x) = \exp(k(x))$, which is in the form required for the generalised trapezoidal rule.
\subsection{Adaptive quadrature} \label{sec:adaptive_quad}
One can also employ a more traditional method of finding the bounce-average, namely by means of adaptive quadrature methods \cite{calvetti2000computation,gander2000adaptive,gonnet2012review}, which furthermore are readily available in a wide assortment of codes. Such methods employ quadrature rules on an iteratively more refined grid, and are able to accurately resolve some integrable singularities. Quadrature methods also tend to outperform naive trapezoidal methods in terms of accuracy, though it should be noted that the computational cost may be higher (especially when trying to accurately resolve a singularity). \par 
There are a few points one needs to be aware of when implementing quadrature rules for bounce-averaging operations. 
\begin{itemize}
    \item One should use a more advanced interpolation scheme for finding the various functions required for the bounce-average than simple piece-wise linear interpolation. If one does use piece-wise linear interpolation, the previously presented generalisation of the trapezoidal rule gives the exact result. 
    \item If the chosen interpolation rule does not allow finding the exact roots of $f(x)=0$, one should take care when evaluating the integrand in the vicinity of the end points. Suppose we numerically find two solutions of $f(x)=0$, namely $x_0$ and $x_1$, within some tolerance $\delta x$. The integral using the estimated roots then becomes
    \begin{equation}
        \int_{x_0 + \delta x_0}^{x_1 + \delta x_1} \frac{h(x)}{\sqrt{f(x)}} \mathrm{d}x,
    \end{equation}
    which becomes imaginary if $f(x_0+ \delta x_0)<0$ or $f(x_1 + \delta x_1)<0$. There are a couple of ways to circumvent this issue. One could shrink the integration domain by some amount to ensure that $f(x)$ is always positive in the domain. For example, given a tolerance of $\delta x$ for the roots, one could shrink the domain as
    \begin{equation}
        \int_{x_0 + \delta x_0+\sqrt{\delta x}}^{x_1 + \delta x_1-\sqrt{\delta x}} \frac{h(x)}{\sqrt{f(x)}} \mathrm{d}x.
    \end{equation}
    One could also ensure positive definiteness by making certain that $f(x)\geq0$, for example by changing the integral to
    \begin{equation}
        \int_{x_0 + \delta x_0}^{x_1 + \delta x_1} \frac{h(x)}{\sqrt{|f(x)|}} \mathrm{d}x.
    \end{equation}
    Both these suggestions converge to the true domain as $\delta x$ goes to zero, and do not encounter problems with imaginary numbers.
    \item It is not known \emph{a priori} how many roots the function $f(x)$ has if one uses an arbitrary interpolation scheme. Thus, care should be taken to accurately resolve all roots. Such problems may be avoided by using monotonic interpolation methods, where the number of roots is the same as in linear interpolation methods. 
\end{itemize}
Generally, quadrature is preferred in scenarios where there are only few nodes on which the various functions are sampled. A sufficiently advanced interpolation scheme may then be able to accurately reconstruct the underlying functions, which in turn can be integrated with adaptive quadrature. \par 
We finally note that one can also circumvent the singular behaviour by subtracting the singularity as is done by \citet{velasco2020knosos}, and such a method may indeed speed up calculations significantly. It does however require accurate information on the derivatives of $f(x)$, which is not always given. If such information is available however, such a method may be preferable in combination with quadrature.

\subsection{Orbits crossing computational boundaries}
Although the presented methods can be used to accurately calculate the drift given the full magnetic geometry, one is often given various quantities along a magnetic field-line with a truncated domain (e.g. $\zeta \in [\zeta_\mathrm{min},\zeta_\mathrm{max}]$). For example, gyrokinetic flux-tube simulations make use of such truncated domain as an exceedingly long domain would be computationally costly to simulate. In general, there will therefore exist magnetic wells that cross the domain boundary, and we need some way of treating these. One should realise that in the most general case it is impossible to accurately reconstruct the BAD of such particles, as it would require information of the magnetic field outside of the given domain. If the magnetic field is rotationally symmetric, however (as is the case in tokamaks), it is possible to reconstruct the BAD, and we thus consider how ways of approximating the BAD beyond the domain in cases where the field is taken to be almost periodic. \par 
It is important to note that, even if the magnetic field is periodic, not all functions entering into the calculation share this property, i.e., given the field-line-following coordinate $\zeta$ and a total domain size of $\zeta_\mathrm{tot}$, not all functions of relevance satisfy
\begin{equation}
    f(\zeta+N\zeta_\mathrm{tot}) = f(\zeta)
\end{equation}
with $N \in \mathbb{Z}$. For the direct averaging method or geometric methods in Clebsch coordinates this condition is inapplicable in devices with significant shear, as there is a non-periodic function proportional to shear which enters into the projection of the drift. To see this, we consider the drift using the direct averaging method discussed in section \ref{sec:direct-averaging}. The effect of shear can most readily be seen by  equation \eqref{eq:relation-gradandcurv} which relates curvature and gradient drifts to each other,
\begin{equation}
    \frac{\mathbf{B} \times \nabla B}{B^2} = \mathbf{b} \times \bm{\kappa} - \mathbf{b} \times \frac{\mu_0 \nabla p}{B^2}.
\end{equation}
If one splits the curvature into a geodesic and a normal part, $\bm{\kappa} = \kappa_n \mathbf{n} + \kappa_g \mathbf{b} \times \mathbf{n} $, one can write this expression in a more convenient form. We find
\begin{equation}
    \frac{\mathbf{B} \times \nabla B}{B^2} = \left( \kappa_n -  \frac{\mu_0 |\nabla p|}{B^2}  \right)  \mathbf{b} \times \mathbf{n} - \kappa_g \mathbf{n},
\end{equation}
and one can readily take the inner product with the binormal vector $\nabla \alpha$. This results in
\begin{equation}
    \frac{\mathbf{B} \times \nabla B}{B^2} \cdot \nabla \alpha = \left( \kappa_n - \kappa_g \cot \vartheta_s - \frac{\mu_0 |\nabla p|}{B^2} \right) \frac{B}{|\nabla \psi|},
\end{equation}
where $\cot\vartheta_s$ is the cotangent of the angle between the $\nabla \psi$ and $\nabla \alpha$ vectors, i.e. $\cot\vartheta_s = (\nabla \psi \cdot \nabla \alpha)/| \nabla \psi \times \nabla \alpha |$. In a tokamak, we see that that the functions, $B$, $\nabla \psi$, $\kappa_n$, and $\kappa_g$ are all periodic. Only $\cot\vartheta_s$ is not periodic, unless the global magnetic shear $s$ is exactly zero. Note that the radial projection
\begin{equation}
    \frac{\mathbf{B} \times \nabla B}{B^2} \cdot \nabla \psi = -\kappa_g |\nabla \psi|
\end{equation}
\emph{does} contain periodic functions only. To now accurately reconstruct the drift beyond the boundary, we make use of the following relation \cite{hegna2000local}
\begin{equation}
    \cot \vartheta_s(\zeta) =-\frac{|\nabla \psi|^2}{B} \int_{\varphi_0}^{\varphi} \left( \frac{\mathrm{d} \iota}{\mathrm{d} \psi} + \frac{\partial D}{\partial \varphi} \right) \mathrm{d} \varphi,
    \label{eq:hegna_shear}
\end{equation}
where $\varphi_0$ defines the centre point of the field line, and $D$ is related to the local magnetic shear via equation \eqref{local3D_D_and_sloc}. From equation \eqref{eq:hegna_shear} one can show that in case of a tokamak, $\cot \vartheta_s$ consists of the sum of a secular term proportional to shear and a periodic term. The function thus behaves as a so-called arithmetic quasi-periodic function \cite{corduneanu2009almost,oka2022space}. Physically, the secular term corresponds to the magnetic shear having the same effect on the angle between $\nabla \psi$ and $\nabla \alpha$ each traversal around the torus, i.e. the same shear mapping is simply applied $N$ times for $N$ traversals. From equation \eqref{eq:hegna_shear} we can now choose a natural boundary condition if we choose the field-line following coordinate to be $\zeta=\varphi$ and $\zeta_\mathrm{tot}=2\pi$, namely we impose that $D(\varphi+N 2 \pi)=D(\varphi)$ as in the previous case, and thus
\begin{equation}
    \cot \vartheta_s(\varphi) = -\frac{|\nabla \psi|^2}{B} \left( \frac{\mathrm{d} \iota}{\mathrm{d} \psi} (\varphi - \varphi_0) + \int_{\varphi_0}^{\varphi} \frac{\partial D}{\partial \varphi} \mathrm{d} \varphi \right).
\end{equation}
Henceforth we shall refer to this behavior as quasi-periodic.
\par 
There are methods in which quasi-periodicity need not be enforced. This can be be achieved by choosing an analytical framework of calculating the bounce-averaged drift in which there are no non-periodic terms, for example by expressing $\partial_\psi\mathcal{J}$ in terms of straight-field-line coordinates \cite{hegna2015effect}. One can also avoid the issue altogether by choosing the domain in such a way that the absolute maxima of the magnetic field strength lie at the boundaries, $B(\zeta_\mathrm{min})=B(\zeta_\mathrm{max})=B_\mathrm{max}$, so that there are no magnetic wells or trapped particles that cross the boundaries.
\section{Examples, comparisons, and results}
In this section, we compare the various numerical and analytical methods against one another, and we use the presented framework on an example. To do so, we investigate three examples: an analytical magnetic well, a large-aspect-ratio tokamak with magnetic shear and a pressure gradient, and a stellarator equilibrium.
\subsection{Simple parabolic well}
\begin{figure}
    \centering
    \includegraphics[width=0.4\textwidth]{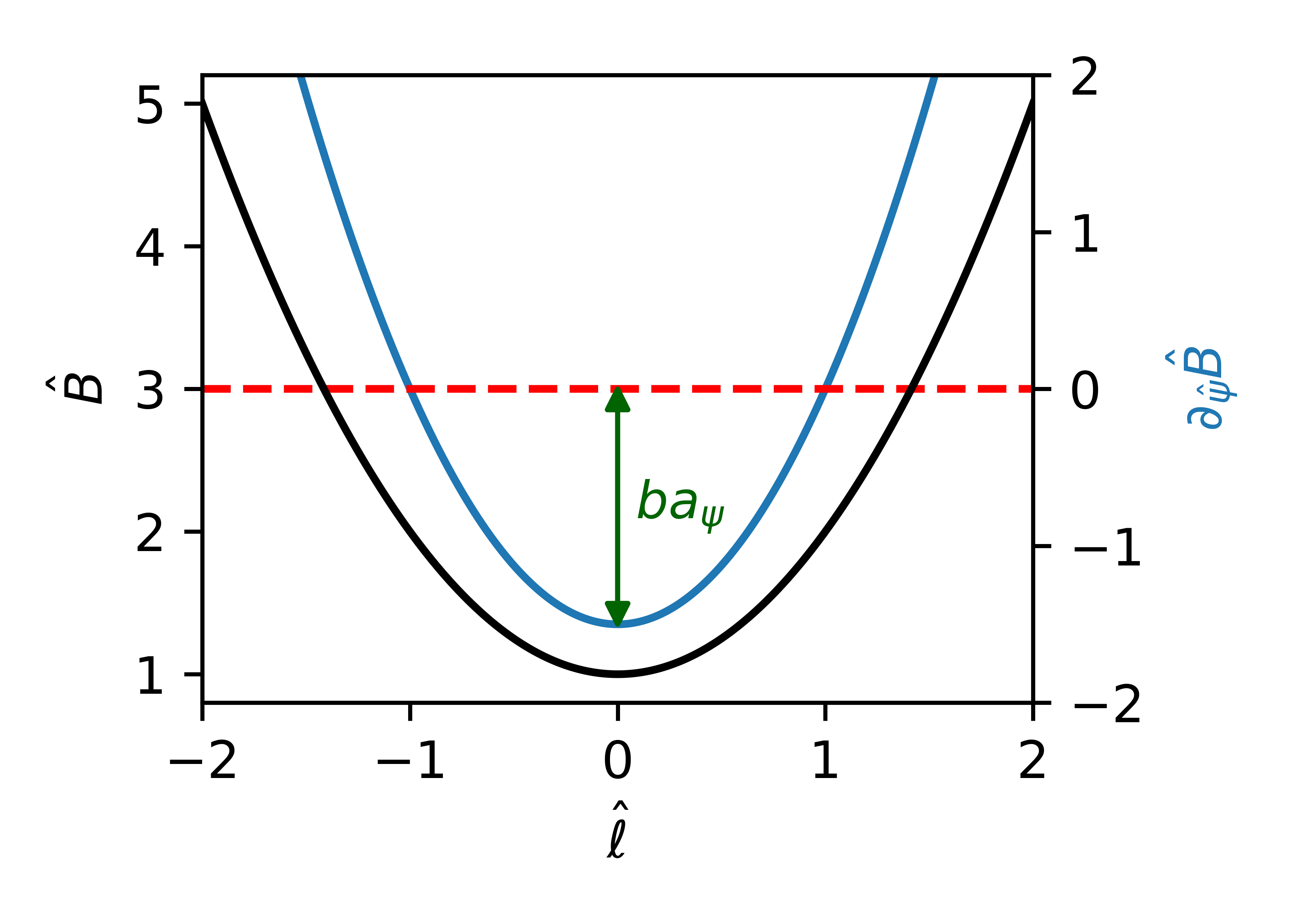}
    \caption{Plot showcasing the magnetic well, and the radial derivative of the magnetic field strength. $b a_\psi$ parameterizes the depth of the well . We use $a_\psi = 3/2$ and $ a_\ell = b = 1$.}
    \label{fig:square_well_plot}
\end{figure}

Consider the following magnetic well
\begin{equation}
    B(\psi,\alpha,\ell) = B_0 \left(1 + a_\ell \hat{\ell}^2  + a_\psi \hat{\psi} \left[ \hat{\ell}^2 - b \right] + \alpha \hat{\ell} \right),
\end{equation}
where $\hat{\ell} = \ell / \ell_0$, $\hat{\psi}=\psi/\psi_0$, $a_\ell$ and $b$ are some positive constants, and $a_\psi$ is an arbitrary constant. We have chosen $\hat{\psi}=\alpha=0$ to represent the field line on which we would like to calculate the bounce-averaged drift. We furthermore neglect the effect of any potential, i.e. we set $\Phi = 0$. A plot showing this well geometry is displayed in Fig. \ref{fig:square_well_plot}.
This problem lends itself most readily to the geometric approach of calculating the BADs. To do so, let us first write down the second adiabatic invariant
\begin{equation}
    \mathcal{J} = \sqrt{2mH} \int \mathrm{d} \ell \sqrt{1 - \lambda \hat{B} }.
\end{equation}
To find the bounce points, we find the longitudinal coordinates where the argument of the square root vanishes, for $\psi=\alpha=0$. This is a linear equation in $\hat{\ell}^2$, and it's solution is
\begin{equation}
    \hat{\ell}_s^2 = \frac{1-\lambda}{a_\ell \lambda}.
\end{equation}
It should be evident that solutions only exist if $\lambda \in [0,1]$. Let us now calculate the various components required for the BADs. We first focus on the bounce time, which can be found by means of equation \eqref{subeq_dJ:3}. We find that it reduces to 
\begin{equation}
    \tau_b = \sqrt{\frac{m}{2H}} \int_{-\ell_s}^{+\ell_s} \frac{\mathrm{d} \ell}{\sqrt{1 - \lambda \hat{B}}}.
    \label{eq:bounce-time}
\end{equation}
This integral is readily evaluated at $\hat{\psi}=\hat{\alpha}=0$, and we find that it reduces to
\begin{equation}
    \tau_b = \ell_0 \sqrt{\frac{m}{2H}} \frac{\pi}{\sqrt{a_\ell \lambda}}.
\end{equation}
Our next step is to evaluate the binormal drift. From equation \eqref{subeq_dJ:1} we have that
\begin{equation}
    \Delta \alpha = \frac{\ell_0}{q \psi_0 }\sqrt{\frac{mH}{2}} \int \frac{ \lambda \partial_{\hat{\psi}} \hat{B}}{\sqrt{1 - \lambda \hat{B}}} \mathrm{d} \hat{\ell}
    \label{eq:Delta_alpha_integral}
\end{equation}
This again can be evaluated at $\hat{\psi}=\alpha=0$, and we find
\begin{equation}
    \Delta \alpha = - \frac{\ell_0}{q \psi_0 }\sqrt{\frac{mH}{2}} \frac{\pi a_\psi (\lambda + 2 \lambda b a_\ell - 1)}{2 a_\ell^{3/2} \sqrt{\lambda}}.
\end{equation}
Finally, since $\partial_\alpha B$ is an even function in $\hat{\ell}$, the radial excursion of \eqref{subeq_dJ:2} is zero. Thus we have that the binormal BAD becomes
\begin{equation}
\langle \mathbf{v}_D \cdot \nabla \alpha \rangle = - \frac{H}{q \psi_0} \frac{a_\psi \left( \lambda + 2 \lambda b a_\ell - 1 \right)}{2 a_\ell}
\end{equation}
Finally, we recast the above equation in terms of the trapping parameter $k^2$ which relates to the pitch angle $\lambda$ as
\begin{equation}
    k^2 = 1 - \lambda,
\end{equation}
so that $k^2=0$ is deeply trapped particles and $k^2=1$ is shallowly trapped particles. It should be noted that this definition of $k^2$ is only valid in the current parabolic well. Using this parameterisation of trapped particles we find that
\begin{equation}
\langle \mathbf{v}_D \cdot \nabla \alpha \rangle = \frac{H}{q \psi_0} \frac{a_\psi \left( k^2 -  2 b a_\ell (1 - k^2) \right)}{2 a_\ell}.
\end{equation}
Deeply trapped particles are drifting as
\begin{equation}
    \lim_{k^2\rightarrow 0} \langle \mathbf{v}_D \cdot \nabla \alpha \rangle = - \frac{H}{q \psi_0} b a_\psi,
\end{equation}
whereas shallowly trapped particles drift as
\begin{equation}
    \lim_{k^2\rightarrow 1} \langle \mathbf{v}_D \cdot \nabla \alpha \rangle = \frac{H}{q \psi_0} \frac{a_\psi}{2 a_\ell}.
\end{equation} 
Thus, as we can see this parabolic well has deeply and shallowly trapped particles drifting in opposite directions, and hence there is always some part of the trapped particle population which destabilises trapped particle modes \cite{rosenbluth1968,proll2012resilience}.
\par 
\begin{figure}
\vspace{5pt}%
\includegraphics[width=0.4\textwidth]{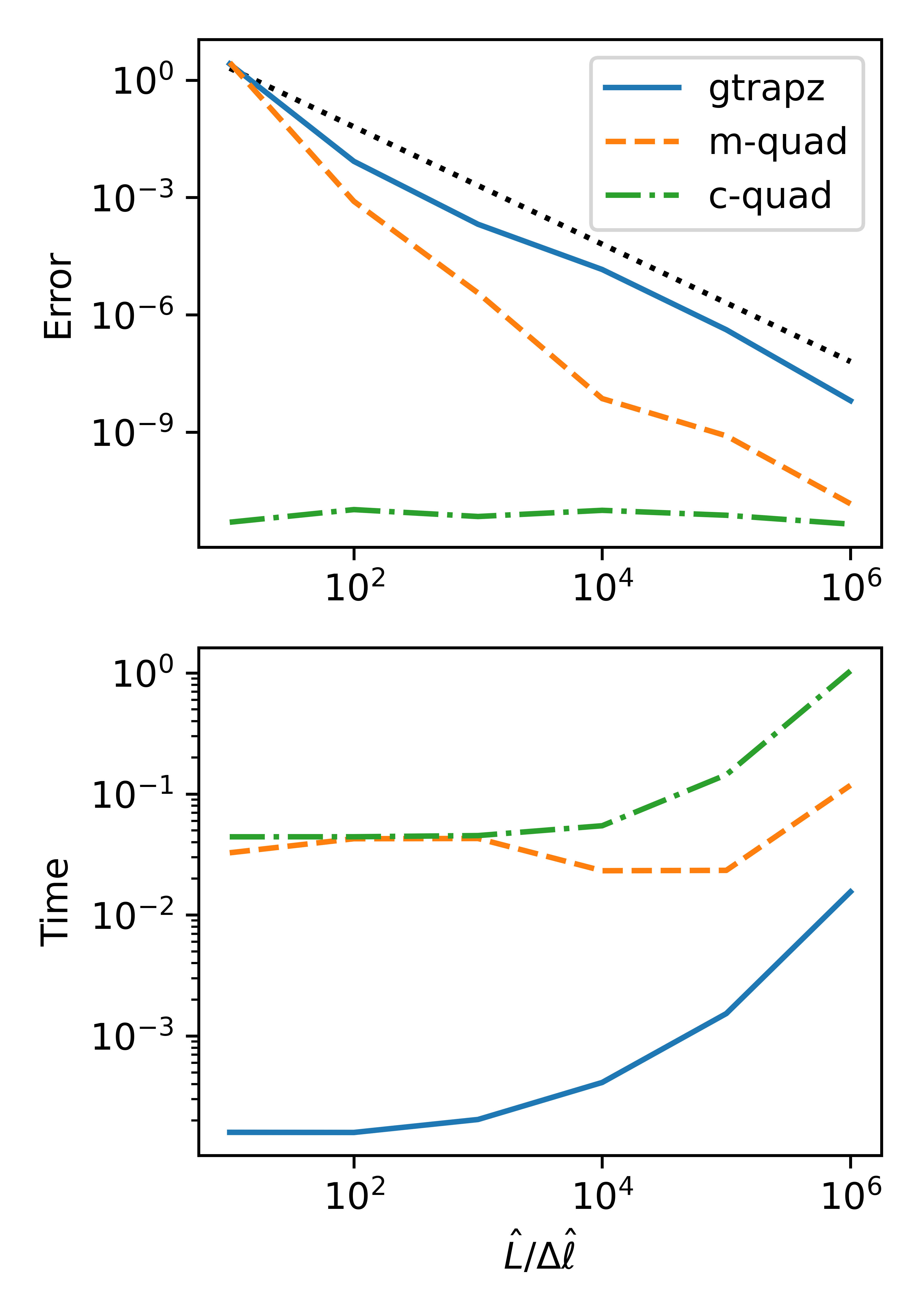}
\caption{\label{fig:err-analysis} The upper plot displays the average relative error for various methods of calculating the bounce-averaged drift, and the black line furthermore shows a $3/2$ scaling-law. The x-axis, $\hat{L}/\Delta \hat{\ell}$, is the total number of samples in the domain. In this plot we have set $a_\psi = 3/2$ and $ = a_\ell = b = 1$. In the bottom plot we show the total average time it takes to calculate one bounce-averaged drift.} 
\end{figure}
We now go on to calculate this BAD and compare numerical methods by computing the integrals of equation \eqref{eq:bounce-time} and \eqref{eq:Delta_alpha_integral} numerically. To compare the various methods, we calculate the relative error in $\langle \mathbf{v}_D \cdot \nabla \alpha \rangle$ for several $k^2$ values, and we average over those. We furthermore investigate the average time it takes to calculate $\langle \mathbf{v}_D \cdot \nabla \alpha \rangle$ for one value of $k^2$. The results can be seen in Fig. \ref{fig:err-analysis}. Several methods are displayed, where gtrapz denotes the generalized trapezoidal method discussed in section \ref{eq:sec-gtrapz}, and two different implementations of adaptive quadrature methods as discussed in section \ref{sec:adaptive_quad}. The two quadrature methods both make use of \texttt{scipy}'s standard adaptive quadrature integration methods to perform the integrals. Their difference is in how the underlying functions (e.g. $B$, $\partial_\psi B$) are interpolated. The m-quad method makes use of monotonic piecewise cubic hermite interpolating polynomials to interpolate the underlying functions \cite{fritsch1984method}, whereas the c-quad method uses cubic splines. \par
From all the displayed methods in Fig. \ref{fig:err-analysis}, we see that the gtrapz method typically performs the worst in terms of accuracy but best in terms of run-time. Its error furthermore scales slightly worse than one would expect from a trapezoidal-like rule, as it has a $3/2$ scaling (displayed as a dotted black line) instead of a square scaling with grid-size. The m-quad method, on the other hand, has a higher accuracy, though at a cost of drastically increased run-time. Since the underlying function cannot be constructed to machine-precision with the m-quad interpolation method, we see that the error decreases with the amount of grid-points. This is in contrast to the c-quad method, which is able to reconstruct the underlying functions using the interpolation method. Accordingly, the error is almost independent of the grid-size, and its value is set by tolerances in both the integration method and the root-finding algorithm. It should be noted that c-quad does have the longest run-time. \par 
We mention in passing that other quadrature methods have also been tested, though these results are not shown here. A particularly good method would be so-called sinh-tanh quadrature rules, as these work well in scenarios where the integrand becomes singular at edge-points\cite{takahasi1974double}. Using this method to perform the integrals does indeed lead to a significant decrease in run-time as compared to the standard quadrature methods of \texttt{scipy}, though the trapezoidal method still tends to be the fastest. We have however found that the result of sinh-tanh rules is strongly dependent on the underlying interpolator, and particularly poor choices can result in this method not converging. \par
All in all, we see that the parabolic well provides a good benchmark for comparing different numerical routines, and these exhibit different degrees of accuracy when the grid-size is varied. The computational cost also varies between the routines, and there is thus no single best method in terms of speed, accuracy, and ease of implementation. Finally, one should also be aware that the interpolation routine can have a significant effect on the error; finding an interpolation method which can accurately reconstruct the underlying function can drastically decrease the error. We finally note that the current results derive from a python implementation. An analogous package has also been constructed in julia, and the results have the same qualitative trends, though the difference in run-time is smaller.

\subsection{Large-aspect-ratio circular tokamak} \label{sec:tokamak-CHM}
In this section, we compare two analytical methods for calculating the BAD, and investigate the BAD of orbits crossing the computational boundary. We do so in the geometry of a large-aspect-ratio tokamak with circular cross-section, magnetic shear, and a pressure gradient. This geometry has previously been investigated in this context by \citet{connor1983effect}, who derived analytical expressions for the bounce-averaged drift exist in this limit. It was found that in this geometry the BAD in the binormal direction becomes
\begin{equation}
    \langle \mathbf{v}_d \cdot \nabla \alpha \rangle = - \frac{2 H}{q B_0 R_0 r} \left( G_1 - \frac{(\iota \alpha)^2}{4} + 2 s G_2 - \alpha G_3  \right),
\end{equation}
where $R_0$ is the major radial coordinate, $r$ is the minor radial coordinate, $\iota$ is the rotational transform, $s = - r \partial_r \ln(\iota)$ is the magnetic shear, and $\alpha = - (2 \mu_0 R_0/ \iota^2 B_0^2) \partial_r p $ is the normalized pressure gradient with $p(r)$ being the pressure. The functions $G_i$ are dependent on the trapping parameter $k^2$, which in general geometries is defined as 
\begin{equation}
    k^2 = \frac{\hat{B}_\mathrm{max} - \lambda \hat{B}_\mathrm{min} \hat{B}_\mathrm{max} }{\hat{B}_\mathrm{max} -  \hat{B}_\mathrm{min} },
    \label{eq:k2-def}
\end{equation}
where the subscripts max and min denote maximal and minimal values of the function. With this trapping parameter, the functions $G_i$ are
\begin{subequations}
\label{eq:whole}
\begin{equation}
G_1(k^2) = \frac{E(k^2)}{K(k^2)} - \frac{1}{2},
\end{equation}
\begin{eqnarray}
G_2(k^2) = \frac{E(k^2)}{K(k^2)} + k^2 - 1,
\end{eqnarray}
\begin{eqnarray}
G_3(k^2) = \frac{2}{3} \left( \frac{E(k^2)}{K(k^2)} (2k^2 - 1) + 1 - k^2 \right),
\end{eqnarray}
\end{subequations}
where $K(k^2)$ and $E(k^2)$ are the complete elliptic integrals of the first and second kind, respectively. One can generate profiles of either the various functions entering $\partial_\psi \mathcal{J}$, or the projections of the drift $\mathbf{v}_D \cdot \nabla x$, by means of the NE3DLE code \cite{JDuff_NE3DLE}. This code generates solutions to the magnetohdyrodynamic equilibrium equations local to a flux surface from a prescribed three-dimensional shaping, $\iota$, and two of either the pressure gradient, global shear, or flux surface averaged parallel current \cite{hegna2000local}. Hence, one can readily prescribe a large-aspect ratio circular flux-surface with shear and a pressure gradient, in line with the assumptions used by \citet{connor1983effect}.
In this geometry we examine differences between analytical methods of calculating the drift. These methods are equivalent, but with the chosen boundary conditions slight differences can show up for particles which cross the boundary. For the numerical integration, we choose the c-quad scheme of the previous section. \par 
\begin{figure}
    \centering
    \includegraphics[width=0.4\textwidth]{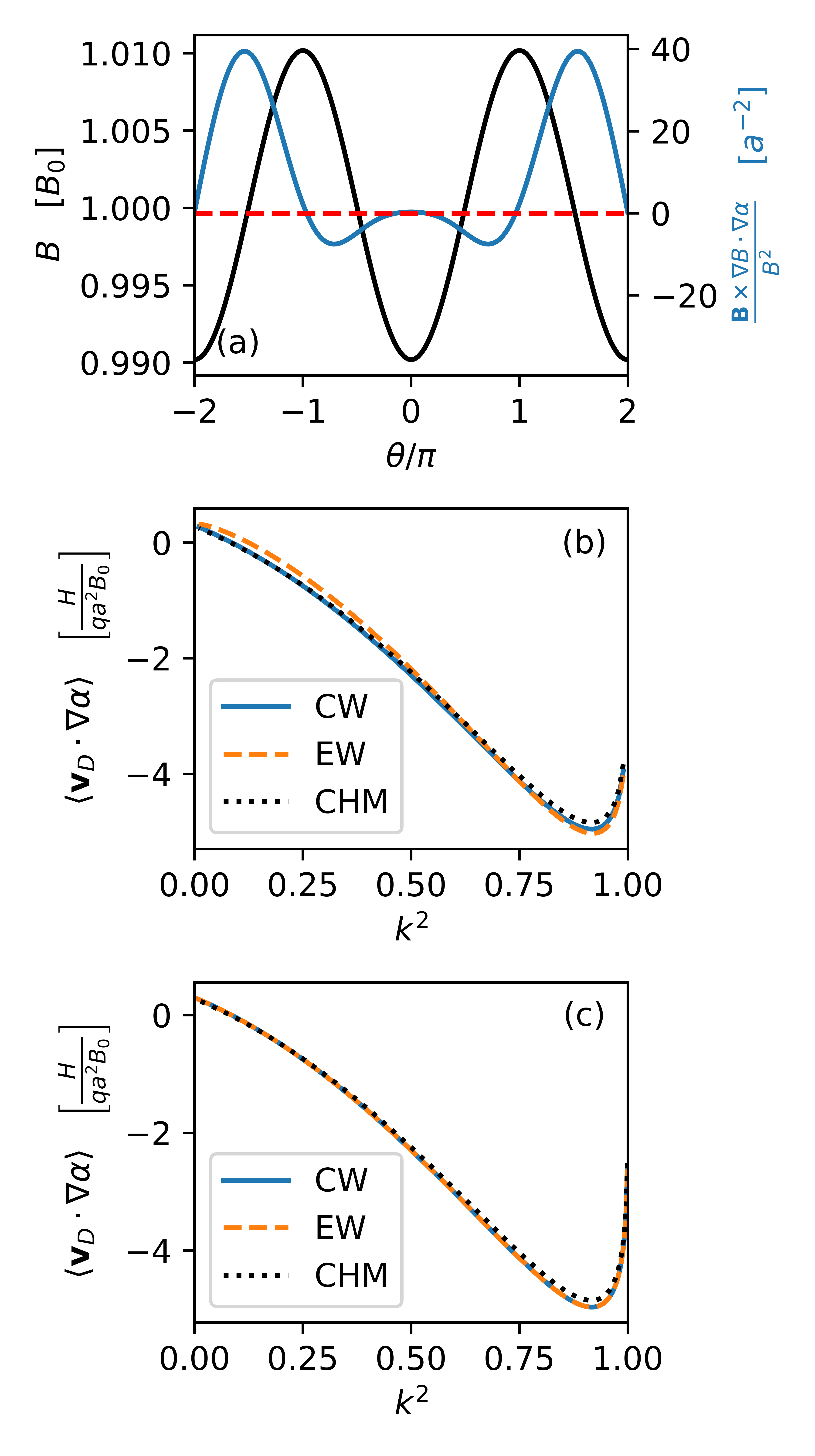}
    \caption{Plot (a) shows magnetic field strength and the gradient drift. In plot (b) and (c) we see the bounce-averaged drift as calculated via the direct averaging method as a function of the trapping parameter $k^2$. Plot (b) uses the quasi-periodic boundary condition, and (c) displays the result as calculated with the geometric method. CHM is the analytical result, CW is the drift in the central well, EW is the drift in the well which crosses the boundary.}
    \label{fig:CHM-direct-averaging}
\end{figure}
The results are plotted in Fig. \ref{fig:CHM-direct-averaging}, where we have chosen $s=7$ and $\alpha = 5$. In plot (a) the magnetic field strength is displayed together with the gradient drift in blue, as a function of the poloidal angle $\theta$. The plots (b) and (c) show the BAD calculated using the direct averaging method from section \ref{sec:direct-averaging} and the geometric method discussed in section \ref{sec:geom-approach} respectively. The latter was done in straight-field-line coordinates, in which quasi-periodicity need not be enforced. \par 
We focus on plot (b) first, where one can see that the analytical result (denoted as CHM) and numerical result for the central well (denoted as CW) line up to a great degree. The well which crosses the computational boundary, denoted as EW (edge well, i.e. $\theta \in (-2\pi,-\pi] \cup [\pi,2\pi)$), has bounce-averaged drifts shown in yellow agree to a great extent, though the match is not perfect. Finally in plot (c) the match between the edge well and central well is exact. This is a consequence of the geometric method used to calculate the drift; the functions entering into the $\partial_\psi\mathcal{J}$ are all exactly periodic and the boundary condition is thus met. \par 
One should realize that all the boundary conditions chosen here work only in special cases. If one were to arbitrarily choose the domain $\theta \in \left[ \theta_\mathrm{min},\theta_\mathrm{max} \right]$ the underlying function are not guaranteed to be (quasi-)periodic in that domain. The same holds true for arbitrary magnetic field-lines on three-dimensional equilibria, as the periodicity of the field-line is then broken. 
\subsection{Stellarator equilibrium and the effect of a radial electric field}
In this section, we calculate the drift in a general three-dimensional equilibrium. We choose to investigate the NCSX equilibrium \cite{zarnstorff2001physics}, as both radial and binormal drifts exist in this configuration. We furthermore artificially add a radial electric field to the system, and we investigate how this changes the trapped particle drifts. \par 
To investigate how a radial electric field changes the BADs, we use the direct averaging approach discussed in section \ref{sec:direct-averaging}. We suppose there exists a radial electric field of the form
\begin{equation}
    \mathbf{E} = E_\psi \nabla \psi.
\end{equation}
It should be evident that the radial electric field only contributes to the binormal drift, as $\mathbf{E} \times \mathbf{B} \cdot \nabla \psi = 0$. This binormal drift is readily found to be constant,
\begin{equation}
    \frac{\mathbf{E} \times \mathbf{B}}{B^2} \cdot \nabla \alpha = E_\psi.
\end{equation}
Since the bounce-averaging operator is linear, the total BAD is found as a sum of the individual components. As such the binormal drift becomes
\begin{equation}
\begin{aligned}
    \langle \mathbf{v}_D \cdot \nabla \alpha \rangle &= E_\psi + \frac{H}{q B_0} \times \\
    &  \left\langle \left(  \lambda \frac{\mathbf{B} \times \nabla B}{B^2} + 2 (\hat{B}^{-1}-\lambda) \frac{\mathbf{B} \times \bm{\kappa}}{B} \right) \cdot \nabla \alpha \right\rangle.
\end{aligned}
\end{equation}
We thus see that the effect of a radial electric-field on bounce-averaged drifts is simple; it simply adds a constant to the drifts. For completeness, we include the radial drift as well which is 
\begin{equation}
\begin{aligned}
    \langle \mathbf{v}_D \cdot \nabla \psi \rangle &=  \frac{H}{q B_0} \times \\
    &  \left\langle \left(  \lambda \frac{\mathbf{B} \times \nabla B}{B^2} + 2 (\hat{B}^{-1}-\lambda) \frac{\mathbf{B} \times \bm{\kappa}}{B} \right) \cdot \nabla \psi \right\rangle.
\end{aligned}
\end{equation}
We go on to note that, although the BAD is affected by the radial electric field, gyrokinetic instabilities in a flux-tube are \emph{unaffected}, except for a simple Doppler shift of the frequency. This may initially seem surprising as it is the sign of the product of the binormal BAD and diamagnetic drifts which determines linear TPM stability. However, the diamagnetic drift wave is also convected along with the $\mathbf{E} \times \mathbf{B}$ drift, and as such one can make a Galilean transformation $\mathbf{v}' = \mathbf{v} - \mathbf{v}_{\mathbf{E} \times \mathbf{B}}$ to a frame in which the effect of the radial electric field is gone \cite{antonsen1980kinetic,catto1981generalized,helander2005collisional,helander2013collisionless}. This argument holds for a flux-tube in a stellarator, or a flux-surface in a tokamak, and as such important differences due to a radial electric field can be found if one investigates global effects \cite{villard2002radial,xanthopoulos2016}. On the flux-surface of a stellarator, for example, a significant electric field makes trapped particles precess around the entire flux surface at a higher pace, and hence these particles experience a flux-surface-averaged drift as opposed to the average along a single field-line. If the flux surface average of the radial drift a trapped particle samples is lower than the radial drift on some individual field lines, the neoclassical transport drops significantly. This mechanism is less effective for energetic particles, as the $\mathbf{E} \times \mathbf{B}$-drift does not scale with particle energy, contrary to the curvature and gradient drifts. \par 
\begin{figure}
\includegraphics[width=0.4\textwidth]{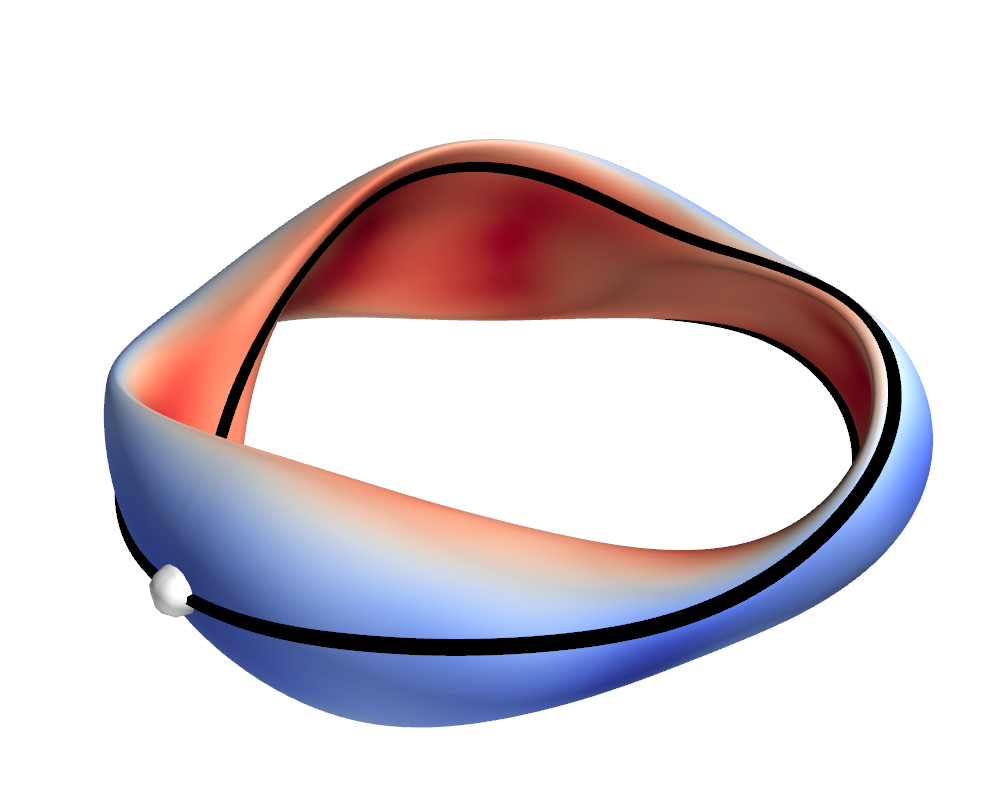}% Here is how to import EPS art
\caption{\label{fig:ncsx-fieldline} The field line in NCSX along which the drifts are calculated is displayed in black, with the white dot indicating the centre point, $\theta=0$. The colour is the magnetic field strength, with red being strong and blue being weak. The surface shown is at $\psi/\psi_\mathrm{edge}=0.5$.} 
\end{figure}
\begin{figure}
\includegraphics[width=0.4\textwidth]{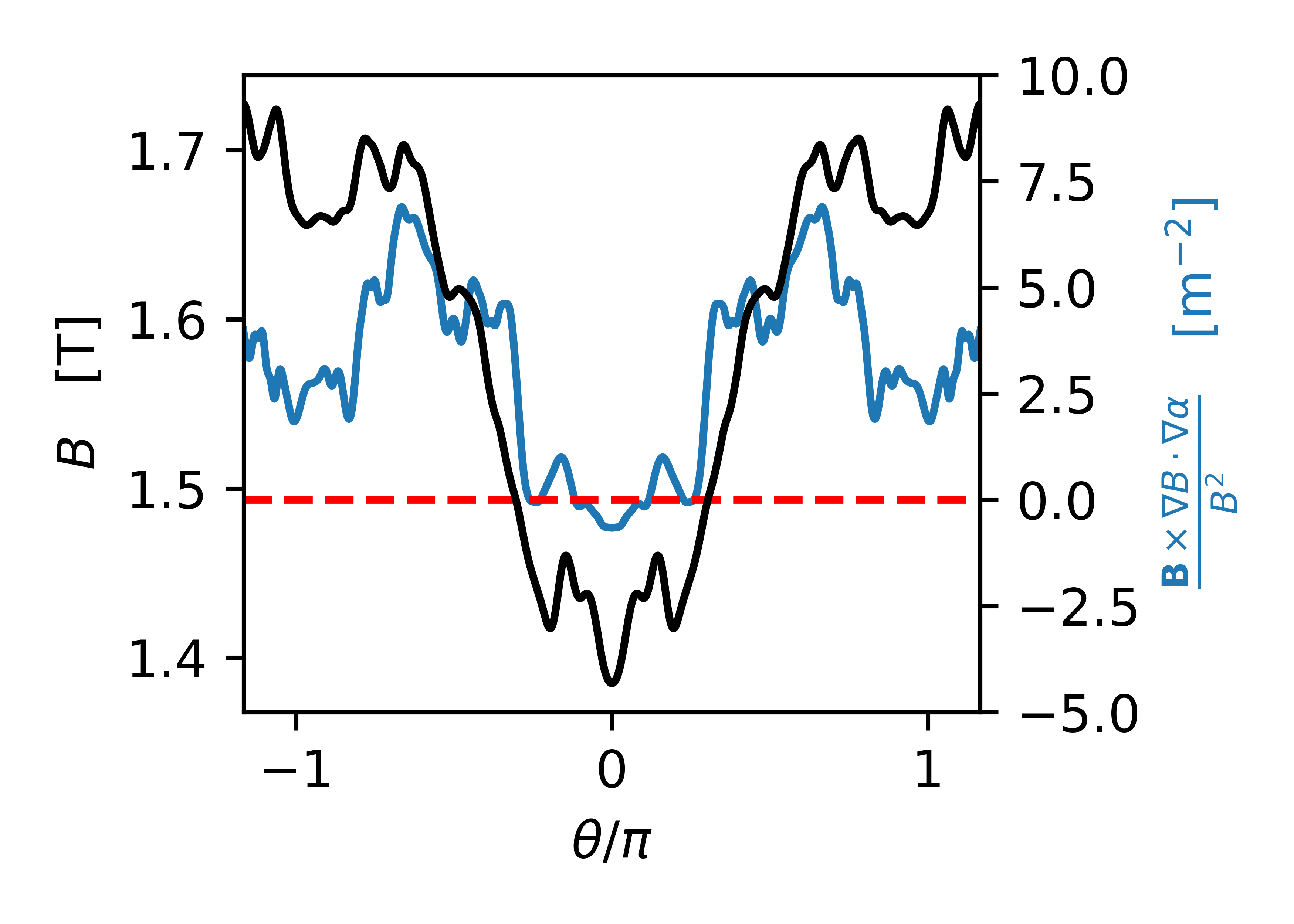}% Here is how to import EPS art
\caption{\label{fig:ncsx-modb} The magnetic field strength along the field line, as a function of the poloidal Boozer angle in black. The blue line is the $\nabla B$-drift, with the red line showing the boundary where the $\nabla B$-drift changes sign.} 
\end{figure}

Let us now investigate the BADs in NCSX without a radial electric field. The field line along which drifts are calculated is shown in Fig. \ref{fig:ncsx-fieldline}, and the magnetic field strength and gradient-drift is shown as a function of the poloidal Boozer angle in Fig. \ref{fig:ncsx-modb}. We have set the domain so that boundary effects are unimportant, by choosing $B(\theta_\mathrm{min})=B(\theta_\mathrm{max})=B_\mathrm{max}$.
To be able to compare the binormal and radial BADs, we first ensure that they have the same units. To do so, we define the radial coordinate in units of length as
\begin{equation}
    r = a \sqrt{\frac{\psi}{\psi_\mathrm{edge}}},
\end{equation}
where $a$ is the minor radius of the device, $\psi$ is the flux surface label, and $\psi_\mathrm{edge}$ is the toroidal flux at the last closed flux surface. We furthermore define a binormal coordinate, also in units of length, as
\begin{equation}
    y = r_0 \alpha,
\end{equation}
where $r_0$ is the radial coordinate of the flux surface in question. Finally, we normalize the drifts as
\begin{subequations}
\label{eq:whole}
\begin{equation}
\langle \hat{\mathbf{v}}_D \cdot \nabla y \rangle = \frac{q a B_0}{H} \frac{\mathrm{d} y}{\mathrm{d} \alpha} \langle \mathbf{v}_D \cdot \nabla \alpha \rangle,\label{normalized-binormal:1}
\end{equation}
\begin{eqnarray}
\langle \hat{\mathbf{v}}_D \cdot \nabla r \rangle =  \frac{q a B_0}{H} \frac{\mathrm{d} r}{\mathrm{d} \psi} \langle \mathbf{v}_D \cdot \nabla \psi \rangle \label{normalized-radial:2}.
\end{eqnarray}
\end{subequations} \par
\begin{figure}
\includegraphics[width=0.44\textwidth]{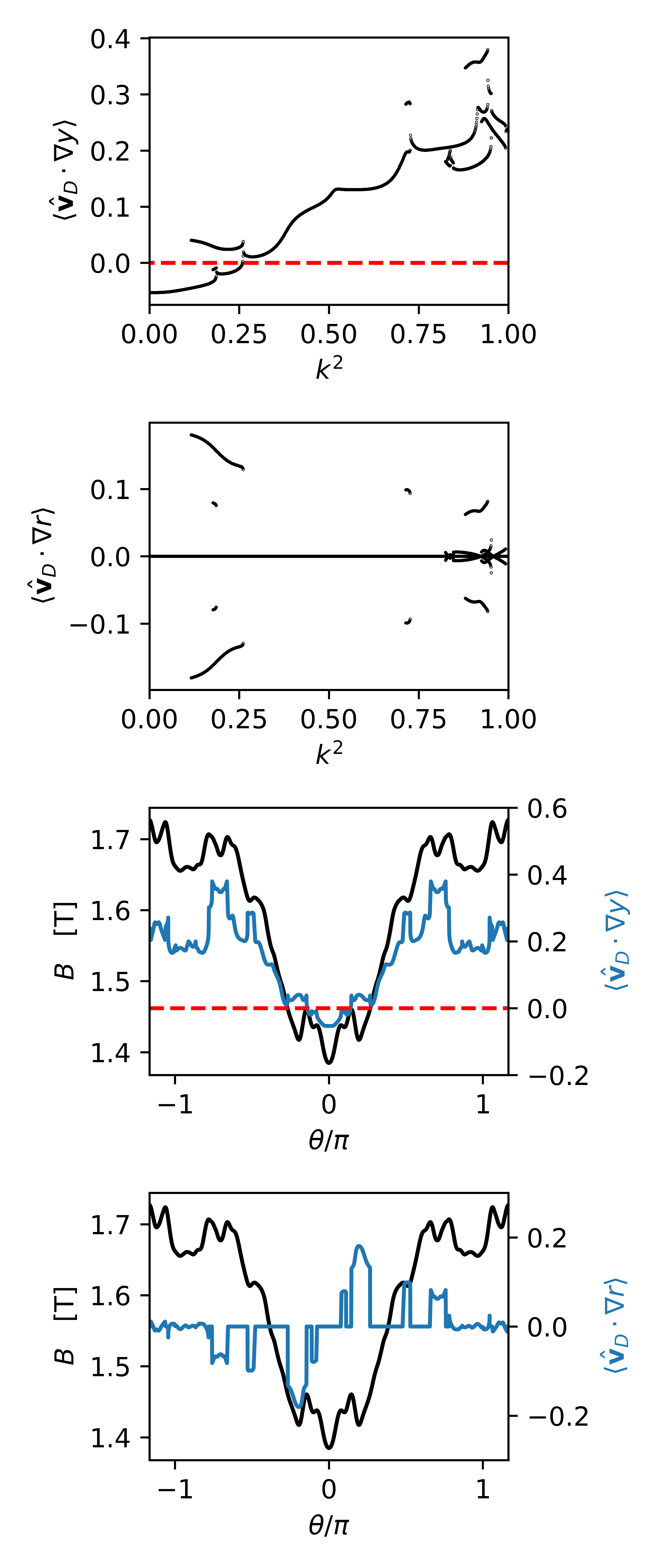}
\caption{\label{fig:ncsx-precession} The top two plots are the binormal and radial drift as a function of $k^2$. The bottom two plots are the same quantities, but as a function of the poloidal bounce-angle.} 
\end{figure}

We plot the normalised drifts as defined in equations \eqref{normalized-binormal:1} and \eqref{normalized-radial:2} in Fig. \ref{fig:ncsx-precession}, as a function of the trapping parameter $k^2$ as in equation \eqref{eq:k2-def}. We furthermore plot the same drifts as a  function of the bounce-angle, that is the $\theta$-value which satisfies the equation 
\begin{equation}
    1-\lambda \hat{B}(\theta) = 0.
\end{equation}
Since one can calculate two bounce-angles for each bounce-well and $\lambda$, which also uniquely define the radial and drifts, one can plot $\langle \mathbf{v}_D \cdot \nabla r \rangle (\theta)$ and $\langle \mathbf{v}_D \cdot \nabla y \rangle (\theta)$.
There are several interesting trends to notice. One first notices multiple discontinuous lines in Fig. \ref{fig:ncsx-precession}, and each line-segment corresponds to a unique bounce-well.  Note that radial drifts typically tend to be smaller than the binormal drifts. This is a consequence of the approximate quasi-axisymmetry NCSX possesses, as exactly quasi-axisymmetric systems have no bounce-averaged radial drift. The central well even has exactly zero radial drifts. This is due to the chosen field line: since it is a stellarator symmetric point, the instantaneous radial drift is an even function in $\theta$ and the bounce-average of the central well is identically zero. This also explains the mirror symmetry of the plots with the bounce-averaged radial drift, as the bounce average of an even function over two symmetric wells differs only in its sign. It is interesting to notice that, although the bounce-averaged radial drift tends to be small, there  are two wells with quite significant drifts. These wells correspond to the two trapping wells adjacent to the central well as can be seen in the bottom plot, and as a consequence these wells may perform especially poor in terms of neoclassical transport. Next, note that the binormal drift is minimal at $k^2=0$, which are the most deeply trapped particles. Invoking the the TPM stability criterion \cite{proll2012resilience}, we find that for a density gradient driven TPM only these deeply trapped particles are destabilising\footnote{An important corollary is that the stability criterion is strictly only true in exactly omnigenous systems.}. \par  
All together we see that plots showing the bounce-averaged drift can be useful in assessing various properties of a magnetic field line, such as finding regions where neoclassical transport is expected to be bad, finding trapped particle populations which are TPM unstable, and assessing the result of an optimisation scheme.
\section{Conclusions}
We have described three methods for calculating the averaged drift of trapped particles: the particle orbit tracing method, the direct averaging method, and the geometric method. Though these are equivalent in the limit of small $\rho_*$, they may vary in ease of calculation in both numerical and analytical scenarios. In order to calculate the bounce-averaged drift in general scenarios we present two numerical methods via which one may solve bounce-averaged integrals, including a generalisation of the trapezoidal rule which works for any integrable singularity, and a more standard quadrature method. \par
We compare performance of various numerical methods on a simple magnetic geometry, namely that of a parabolic trapping well. We find that the generalized trapezoidal rule has an error which decreases with a $3/2$ power with the number of grid-points. The quadature methods on the other hand have error scalings which depend strongly on the interpolation method and chosen numerical tolerances. Generally we find that the generalized trapezoidal rule performs much better in terms of computational cost as compared to a quadrature method, and as such this method may be useful in scanning over large databases or in optimisation loops. \par 
We have also investigated the bounce-averaged drift in the case of a large aspect-ratio tokamak with magnetic shear and a pressure gradient. In this setting we compare different analytical methods of calculating the bounce-averaged drift, and we test the quasi-periodic condition for trapped particles which cross the computational domain. It is found that both the direct averaging method and the geometric method in straight-field-line angular coordinates agree. \par 
Finally, we use the presented framework to investigate bounce-averaged drifts in NCSX. We find that the radial drifts are typically small, which is a consequence of the approximate quasi-axisymmetry NCSX possesses. One can furthermore readily assess linear stability of the TPM by investigating the binormal drifts. We furthermore investigate how a radial electric field would adjust bounce-averaged drifts and we find that the overall effect is simple; it simply adds a constant to the binormal drift. We discuss the consequence of a radial electric field on the trapped particles orbits and argue that neoclassical transport may be reduced. \par 
All in all, we see that the various methods presented here are effective ways to calculate bounce-averaged drifts, and from these bounce-averaged drifts one can quickly infer important properties. The presented methods can furthermore readily be used to calculate the bounce-averaged drift in a wide assortment of codes, ranging from gyrokinetic to equilibrium codes. The code suite has been made available in python and can be installed via standard methods \cite{BAD_code}. It can furthermore be used to calculate the bounce-average of any function, such as the electrostatic potential, as is required for the TPM dispersion relation \cite{helander2013collisionless}.

\section*{Acknowledgements}
The authors are grateful for fruitful discussions with B. Faber, M.J. Pueschel, E. Rodr\'iguez, A. Goodman, P. Mulholland, P. Costello, M. Morren, and R. Jorge. This work was partly supported by a grant from the Simons Foundation (560651, PH), and this publication is part of the project “Shaping turbulence—building a framework for turbulence optimisation of fusion reactors,” with Project No. \texttt{OCENW.KLEIN.013} of the research program “NWO Open Competition Domain Science” which is financed by the Dutch Research Council (NWO). This work has been carried out within the framework of the EUROfusion Consortium, funded by the European Union via the Euratom Research and Training Program (Grant Agreement No. 101052200—EUROfusion). Views and opinions expressed are however those of the author(s) only and do not necessarily reflect those of the European Union or the European Commission. Neither the European Union nor the European Commission can be held responsible for them.

\appendix

\bibliography{aipsamp}

\begin{thebibliography}{45}%
\makeatletter
\providecommand \@ifxundefined [1]{%
 \@ifx{#1\undefined}
}%
\providecommand \@ifnum [1]{%
 \ifnum #1\expandafter \@firstoftwo
 \else \expandafter \@secondoftwo
 \fi
}%
\providecommand \@ifx [1]{%
 \ifx #1\expandafter \@firstoftwo
 \else \expandafter \@secondoftwo
 \fi
}%
\providecommand \natexlab [1]{#1}%
\providecommand \enquote  [1]{``#1''}%
\providecommand \bibnamefont  [1]{#1}%
\providecommand \bibfnamefont [1]{#1}%
\providecommand \citenamefont [1]{#1}%
\providecommand \href@noop [0]{\@secondoftwo}%
\providecommand \href [0]{\begingroup \@sanitize@url \@href}%
\providecommand \@href[1]{\@@startlink{#1}\@@href}%
\providecommand \@@href[1]{\endgroup#1\@@endlink}%
\providecommand \@sanitize@url [0]{\catcode `\\12\catcode `\$12\catcode
  `\&12\catcode `\#12\catcode `\^12\catcode `\_12\catcode `\%12\relax}%
\providecommand \@@startlink[1]{}%
\providecommand \@@endlink[0]{}%
\providecommand \url  [0]{\begingroup\@sanitize@url \@url }%
\providecommand \@url [1]{\endgroup\@href {#1}{\urlprefix }}%
\providecommand \urlprefix  [0]{URL }%
\providecommand \Eprint [0]{\href }%
\providecommand \doibase [0]{http://dx.doi.org/}%
\providecommand \selectlanguage [0]{\@gobble}%
\providecommand \bibinfo  [0]{\@secondoftwo}%
\providecommand \bibfield  [0]{\@secondoftwo}%
\providecommand \translation [1]{[#1]}%
\providecommand \BibitemOpen [0]{}%
\providecommand \bibitemStop [0]{}%
\providecommand \bibitemNoStop [0]{.\EOS\space}%
\providecommand \EOS [0]{\spacefactor3000\relax}%
\providecommand \BibitemShut  [1]{\csname bibitem#1\endcsname}%
\let\auto@bib@innerbib\@empty
%</preamble>
\bibitem [{\citenamefont {Boozer}(1983)}]{boozer1983transport}%
  \BibitemOpen
  \bibfield  {author} {\bibinfo {author} {\bibfnamefont {A.~H.}\ \bibnamefont
  {Boozer}},\ }\bibfield  {title} {\enquote {\bibinfo {title} {Transport and
  isomorphic equilibria},}\ }\href@noop {} {\bibfield  {journal} {\bibinfo
  {journal} {The Physics of Fluids}\ }\textbf {\bibinfo {volume} {26}},\
  \bibinfo {pages} {496--499} (\bibinfo {year} {1983})}\BibitemShut {NoStop}%
\bibitem [{\citenamefont {Rodriguez}, \citenamefont {Helander},\ and\
  \citenamefont {Bhattacharjee}(2020)}]{rodriguez2020necessary}%
  \BibitemOpen
  \bibfield  {author} {\bibinfo {author} {\bibfnamefont {E.}~\bibnamefont
  {Rodriguez}}, \bibinfo {author} {\bibfnamefont {P.}~\bibnamefont {Helander}},
  \ and\ \bibinfo {author} {\bibfnamefont {A.}~\bibnamefont {Bhattacharjee}},\
  }\bibfield  {title} {\enquote {\bibinfo {title} {Necessary and sufficient
  conditions for quasisymmetry},}\ }\href@noop {} {\bibfield  {journal}
  {\bibinfo  {journal} {Physics of Plasmas}\ }\textbf {\bibinfo {volume}
  {27}},\ \bibinfo {pages} {062501} (\bibinfo {year} {2020})}\BibitemShut
  {NoStop}%
\bibitem [{\citenamefont {Beidler}\ \emph {et~al.}(2021)\citenamefont
  {Beidler}, \citenamefont {Smith}, \citenamefont {Alonso}, \citenamefont
  {Andreeva}, \citenamefont {Baldzuhn}, \citenamefont {Beurskens},
  \citenamefont {Borchardt}, \citenamefont {Bozhenkov}, \citenamefont
  {Brunner}, \citenamefont {Damm} \emph {et~al.}}]{beidler2021demonstration}%
  \BibitemOpen
  \bibfield  {author} {\bibinfo {author} {\bibfnamefont {C.}~\bibnamefont
  {Beidler}}, \bibinfo {author} {\bibfnamefont {H.}~\bibnamefont {Smith}},
  \bibinfo {author} {\bibfnamefont {A.}~\bibnamefont {Alonso}}, \bibinfo
  {author} {\bibfnamefont {T.}~\bibnamefont {Andreeva}}, \bibinfo {author}
  {\bibfnamefont {J.}~\bibnamefont {Baldzuhn}}, \bibinfo {author}
  {\bibfnamefont {M.}~\bibnamefont {Beurskens}}, \bibinfo {author}
  {\bibfnamefont {M.}~\bibnamefont {Borchardt}}, \bibinfo {author}
  {\bibfnamefont {S.}~\bibnamefont {Bozhenkov}}, \bibinfo {author}
  {\bibfnamefont {K.}~\bibnamefont {Brunner}}, \bibinfo {author} {\bibfnamefont
  {H.}~\bibnamefont {Damm}},  \emph {et~al.},\ }\bibfield  {title} {\enquote
  {\bibinfo {title} {Demonstration of reduced neoclassical energy transport in
  wendelstein 7-x},}\ }\href@noop {} {\bibfield  {journal} {\bibinfo  {journal}
  {Nature}\ }\textbf {\bibinfo {volume} {596}},\ \bibinfo {pages} {221--226}
  (\bibinfo {year} {2021})}\BibitemShut {NoStop}%
\bibitem [{\citenamefont {Landreman}\ and\ \citenamefont
  {Paul}(2022)}]{landreman2022magnetic}%
  \BibitemOpen
  \bibfield  {author} {\bibinfo {author} {\bibfnamefont {M.}~\bibnamefont
  {Landreman}}\ and\ \bibinfo {author} {\bibfnamefont {E.}~\bibnamefont
  {Paul}},\ }\bibfield  {title} {\enquote {\bibinfo {title} {Magnetic fields
  with precise quasisymmetry for plasma confinement},}\ }\href@noop {}
  {\bibfield  {journal} {\bibinfo  {journal} {Physical Review Letters}\
  }\textbf {\bibinfo {volume} {128}},\ \bibinfo {pages} {035001} (\bibinfo
  {year} {2022})}\BibitemShut {NoStop}%
\bibitem [{\citenamefont {Goodman}\ \emph {et~al.}(2022)\citenamefont
  {Goodman}, \citenamefont {Mata}, \citenamefont {Henneberg}, \citenamefont
  {Jorge}, \citenamefont {Landreman}, \citenamefont {Plunk}, \citenamefont
  {Smith}, \citenamefont {Mackenbach},\ and\ \citenamefont
  {Helander}}]{goodman2022constructing}%
  \BibitemOpen
  \bibfield  {author} {\bibinfo {author} {\bibfnamefont {A.}~\bibnamefont
  {Goodman}}, \bibinfo {author} {\bibfnamefont {K.~C.}\ \bibnamefont {Mata}},
  \bibinfo {author} {\bibfnamefont {S.~A.}\ \bibnamefont {Henneberg}}, \bibinfo
  {author} {\bibfnamefont {R.}~\bibnamefont {Jorge}}, \bibinfo {author}
  {\bibfnamefont {M.}~\bibnamefont {Landreman}}, \bibinfo {author}
  {\bibfnamefont {G.}~\bibnamefont {Plunk}}, \bibinfo {author} {\bibfnamefont
  {H.}~\bibnamefont {Smith}}, \bibinfo {author} {\bibfnamefont
  {R.}~\bibnamefont {Mackenbach}}, \ and\ \bibinfo {author} {\bibfnamefont
  {P.}~\bibnamefont {Helander}},\ }\bibfield  {title} {\enquote {\bibinfo
  {title} {Constructing precisely quasi-isodynamic magnetic fields},}\
  }\href@noop {} {\bibfield  {journal} {\bibinfo  {journal} {arXiv preprint
  arXiv:2211.09829}\ } (\bibinfo {year} {2022})}\BibitemShut {NoStop}%
\bibitem [{\citenamefont {Proll}\ \emph {et~al.}(2012)\citenamefont {Proll},
  \citenamefont {Helander}, \citenamefont {Connor},\ and\ \citenamefont
  {Plunk}}]{proll2012resilience}%
  \BibitemOpen
  \bibfield  {author} {\bibinfo {author} {\bibfnamefont {J.~H.~E.}\
  \bibnamefont {Proll}}, \bibinfo {author} {\bibfnamefont {P.}~\bibnamefont
  {Helander}}, \bibinfo {author} {\bibfnamefont {J.~W.}\ \bibnamefont
  {Connor}}, \ and\ \bibinfo {author} {\bibfnamefont {G.}~\bibnamefont
  {Plunk}},\ }\bibfield  {title} {\enquote {\bibinfo {title} {Resilience of
  quasi-isodynamic stellarators against trapped-particle instabilities},}\
  }\href@noop {} {\bibfield  {journal} {\bibinfo  {journal} {Physical Review
  Letters}\ }\textbf {\bibinfo {volume} {108}},\ \bibinfo {pages} {245002}
  (\bibinfo {year} {2012})}\BibitemShut {NoStop}%
\bibitem [{\citenamefont {Helander}, \citenamefont {Proll},\ and\ \citenamefont
  {Plunk}(2013)}]{helander2013collisionless}%
  \BibitemOpen
  \bibfield  {author} {\bibinfo {author} {\bibfnamefont {P.}~\bibnamefont
  {Helander}}, \bibinfo {author} {\bibfnamefont {J.~H.~E.}\ \bibnamefont
  {Proll}}, \ and\ \bibinfo {author} {\bibfnamefont {G.~G.}\ \bibnamefont
  {Plunk}},\ }\bibfield  {title} {\enquote {\bibinfo {title} {Collisionless
  microinstabilities in stellarators. i. analytical theory of trapped-particle
  modes},}\ }\href@noop {} {\bibfield  {journal} {\bibinfo  {journal} {Physics
  of Plasmas}\ }\textbf {\bibinfo {volume} {20}},\ \bibinfo {pages} {122505}
  (\bibinfo {year} {2013})}\BibitemShut {NoStop}%
\bibitem [{\citenamefont {Helander}(2017)}]{helander2017available}%
  \BibitemOpen
  \bibfield  {author} {\bibinfo {author} {\bibfnamefont {P.}~\bibnamefont
  {Helander}},\ }\bibfield  {title} {\enquote {\bibinfo {title} {Available
  energy and ground states of collisionless plasmas},}\ }\href@noop {}
  {\bibfield  {journal} {\bibinfo  {journal} {Journal of Plasma Physics}\
  }\textbf {\bibinfo {volume} {83}},\ \bibinfo {pages} {715830401} (\bibinfo
  {year} {2017})}\BibitemShut {NoStop}%
\bibitem [{\citenamefont {White}\ and\ \citenamefont
  {Chance}(1984)}]{white1984hamiltonian}%
  \BibitemOpen
  \bibfield  {author} {\bibinfo {author} {\bibfnamefont {R.~t.}\ \bibnamefont
  {White}}\ and\ \bibinfo {author} {\bibfnamefont {M.}~\bibnamefont {Chance}},\
  }\bibfield  {title} {\enquote {\bibinfo {title} {Hamiltonian guiding center
  drift orbit calculation for plasmas of arbitrary cross section},}\
  }\href@noop {} {\bibfield  {journal} {\bibinfo  {journal} {The Physics of
  fluids}\ }\textbf {\bibinfo {volume} {27}},\ \bibinfo {pages} {2455--2467}
  (\bibinfo {year} {1984})}\BibitemShut {NoStop}%
\bibitem [{\citenamefont {Roach}, \citenamefont {Connor},\ and\ \citenamefont
  {Janjua}(1995)}]{roach1995trapped}%
  \BibitemOpen
  \bibfield  {author} {\bibinfo {author} {\bibfnamefont {C.}~\bibnamefont
  {Roach}}, \bibinfo {author} {\bibfnamefont {J.}~\bibnamefont {Connor}}, \
  and\ \bibinfo {author} {\bibfnamefont {S.}~\bibnamefont {Janjua}},\
  }\bibfield  {title} {\enquote {\bibinfo {title} {Trapped particle precession
  in advanced tokamaks},}\ }\href@noop {} {\bibfield  {journal} {\bibinfo
  {journal} {Plasma physics and controlled fusion}\ }\textbf {\bibinfo {volume}
  {37}},\ \bibinfo {pages} {679} (\bibinfo {year} {1995})}\BibitemShut
  {NoStop}%
\bibitem [{\citenamefont {Marinoni}\ \emph {et~al.}(2009)\citenamefont
  {Marinoni}, \citenamefont {Brunner}, \citenamefont {Camenen}, \citenamefont
  {Coda}, \citenamefont {Graves}, \citenamefont {Lapillonne}, \citenamefont
  {Pochelon}, \citenamefont {Sauter},\ and\ \citenamefont
  {Villard}}]{marinoni2009effect}%
  \BibitemOpen
  \bibfield  {author} {\bibinfo {author} {\bibfnamefont {A.}~\bibnamefont
  {Marinoni}}, \bibinfo {author} {\bibfnamefont {S.}~\bibnamefont {Brunner}},
  \bibinfo {author} {\bibfnamefont {Y.}~\bibnamefont {Camenen}}, \bibinfo
  {author} {\bibfnamefont {S.}~\bibnamefont {Coda}}, \bibinfo {author}
  {\bibfnamefont {J.}~\bibnamefont {Graves}}, \bibinfo {author} {\bibfnamefont
  {X.}~\bibnamefont {Lapillonne}}, \bibinfo {author} {\bibfnamefont
  {A.}~\bibnamefont {Pochelon}}, \bibinfo {author} {\bibfnamefont
  {O.}~\bibnamefont {Sauter}}, \ and\ \bibinfo {author} {\bibfnamefont
  {L.}~\bibnamefont {Villard}},\ }\bibfield  {title} {\enquote {\bibinfo
  {title} {The effect of plasma triangularity on turbulent transport: modeling
  tcv experiments by linear and non-linear gyrokinetic simulations},}\
  }\href@noop {} {\bibfield  {journal} {\bibinfo  {journal} {Plasma Physics and
  Controlled Fusion}\ }\textbf {\bibinfo {volume} {51}},\ \bibinfo {pages}
  {055016} (\bibinfo {year} {2009})}\BibitemShut {NoStop}%
\bibitem [{\citenamefont {Proll}\ \emph {et~al.}(2015)\citenamefont {Proll},
  \citenamefont {Mynick}, \citenamefont {Xanthopoulos}, \citenamefont
  {Lazerson},\ and\ \citenamefont {Faber}}]{proll2015tem}%
  \BibitemOpen
  \bibfield  {author} {\bibinfo {author} {\bibfnamefont {J.}~\bibnamefont
  {Proll}}, \bibinfo {author} {\bibfnamefont {H.}~\bibnamefont {Mynick}},
  \bibinfo {author} {\bibfnamefont {P.}~\bibnamefont {Xanthopoulos}}, \bibinfo
  {author} {\bibfnamefont {S.}~\bibnamefont {Lazerson}}, \ and\ \bibinfo
  {author} {\bibfnamefont {B.}~\bibnamefont {Faber}},\ }\bibfield  {title}
  {\enquote {\bibinfo {title} {Tem turbulence optimisation in stellarators},}\
  }\href@noop {} {\bibfield  {journal} {\bibinfo  {journal} {Plasma Physics and
  Controlled Fusion}\ }\textbf {\bibinfo {volume} {58}},\ \bibinfo {pages}
  {014006} (\bibinfo {year} {2015})}\BibitemShut {NoStop}%
\bibitem [{\citenamefont {Gan}\ and\ \citenamefont {Zhou}(2021)}]{gan2021self}%
  \BibitemOpen
  \bibfield  {author} {\bibinfo {author} {\bibfnamefont {C.}~\bibnamefont
  {Gan}}\ and\ \bibinfo {author} {\bibfnamefont {D.}~\bibnamefont {Zhou}},\
  }\bibfield  {title} {\enquote {\bibinfo {title} {A self-consistent local
  equilibrium model and its application to calculation of the trapped electron
  precession},}\ }\href@noop {} {\bibfield  {journal} {\bibinfo  {journal}
  {Physica Scripta}\ }\textbf {\bibinfo {volume} {96}},\ \bibinfo {pages}
  {105603} (\bibinfo {year} {2021})}\BibitemShut {NoStop}%
\bibitem [{\citenamefont {Stephens}\ \emph {et~al.}(2021)\citenamefont
  {Stephens}, \citenamefont {Garbet}, \citenamefont {Citrin}, \citenamefont
  {Bourdelle}, \citenamefont {van~de Plassche},\ and\ \citenamefont
  {Jenko}}]{stephens2021quasilinear}%
  \BibitemOpen
  \bibfield  {author} {\bibinfo {author} {\bibfnamefont {C.~D.}\ \bibnamefont
  {Stephens}}, \bibinfo {author} {\bibfnamefont {X.}~\bibnamefont {Garbet}},
  \bibinfo {author} {\bibfnamefont {J.}~\bibnamefont {Citrin}}, \bibinfo
  {author} {\bibfnamefont {C.}~\bibnamefont {Bourdelle}}, \bibinfo {author}
  {\bibfnamefont {K.~L.}\ \bibnamefont {van~de Plassche}}, \ and\ \bibinfo
  {author} {\bibfnamefont {F.}~\bibnamefont {Jenko}},\ }\bibfield  {title}
  {\enquote {\bibinfo {title} {Quasilinear gyrokinetic theory: a derivation of
  qualikiz},}\ }\href@noop {} {\bibfield  {journal} {\bibinfo  {journal}
  {Journal of Plasma Physics}\ }\textbf {\bibinfo {volume} {87}},\ \bibinfo
  {pages} {905870409} (\bibinfo {year} {2021})}\BibitemShut {NoStop}%
\bibitem [{\citenamefont {Mackenbach}, \citenamefont {Proll},\ and\
  \citenamefont {Helander}(2022)}]{mackenbach2022available}%
  \BibitemOpen
  \bibfield  {author} {\bibinfo {author} {\bibfnamefont {R.}~\bibnamefont
  {Mackenbach}}, \bibinfo {author} {\bibfnamefont {J.~H.}\ \bibnamefont
  {Proll}}, \ and\ \bibinfo {author} {\bibfnamefont {P.}~\bibnamefont
  {Helander}},\ }\bibfield  {title} {\enquote {\bibinfo {title} {Available
  energy of trapped electrons and its relation to turbulent transport},}\
  }\href@noop {} {\bibfield  {journal} {\bibinfo  {journal} {Physical Review
  Letters}\ }\textbf {\bibinfo {volume} {128}},\ \bibinfo {pages} {175001}
  (\bibinfo {year} {2022})}\BibitemShut {NoStop}%
\bibitem [{\citenamefont {Helander}\ and\ \citenamefont
  {Sigmar}(2005)}]{helander2005collisional}%
  \BibitemOpen
  \bibfield  {author} {\bibinfo {author} {\bibfnamefont {P.}~\bibnamefont
  {Helander}}\ and\ \bibinfo {author} {\bibfnamefont {D.~J.}\ \bibnamefont
  {Sigmar}},\ }\href@noop {} {\emph {\bibinfo {title} {Collisional transport in
  magnetized plasmas}}},\ Vol.~\bibinfo {volume} {4}\ (\bibinfo  {publisher}
  {Cambridge university press},\ \bibinfo {year} {2005})\BibitemShut {NoStop}%
\bibitem [{\citenamefont {Hegna}(2015)}]{hegna2015effect}%
  \BibitemOpen
  \bibfield  {author} {\bibinfo {author} {\bibfnamefont {C.}~\bibnamefont
  {Hegna}},\ }\bibfield  {title} {\enquote {\bibinfo {title} {The effect of
  three-dimensional fields on bounce averaged particle drifts in a tokamak},}\
  }\href@noop {} {\bibfield  {journal} {\bibinfo  {journal} {Physics of
  Plasmas}\ }\textbf {\bibinfo {volume} {22}},\ \bibinfo {pages} {072510}
  (\bibinfo {year} {2015})}\BibitemShut {NoStop}%
\bibitem [{\citenamefont {Helander}(2014)}]{helander2014theory}%
  \BibitemOpen
  \bibfield  {author} {\bibinfo {author} {\bibfnamefont {P.}~\bibnamefont
  {Helander}},\ }\bibfield  {title} {\enquote {\bibinfo {title} {Theory of
  plasma confinement in non-axisymmetric magnetic fields},}\ }\href@noop {}
  {\bibfield  {journal} {\bibinfo  {journal} {Reports on Progress in Physics}\
  }\textbf {\bibinfo {volume} {77}},\ \bibinfo {pages} {087001} (\bibinfo
  {year} {2014})}\BibitemShut {NoStop}%
\bibitem [{\citenamefont {Blank}(2004)}]{blank2004guiding}%
  \BibitemOpen
  \bibfield  {author} {\bibinfo {author} {\bibfnamefont {H.~d.}\ \bibnamefont
  {Blank}},\ }\bibfield  {title} {\enquote {\bibinfo {title} {Guiding center
  motion},}\ }\href@noop {} {\bibfield  {journal} {\bibinfo  {journal} {Fusion
  science and technology}\ }\textbf {\bibinfo {volume} {45}},\ \bibinfo {pages}
  {47--54} (\bibinfo {year} {2004})}\BibitemShut {NoStop}%
\bibitem [{\citenamefont {Albert}, \citenamefont {Kasilov},\ and\ \citenamefont
  {Kernbichler}(2020)}]{albert2020accelerated}%
  \BibitemOpen
  \bibfield  {author} {\bibinfo {author} {\bibfnamefont {C.~G.}\ \bibnamefont
  {Albert}}, \bibinfo {author} {\bibfnamefont {S.~V.}\ \bibnamefont {Kasilov}},
  \ and\ \bibinfo {author} {\bibfnamefont {W.}~\bibnamefont {Kernbichler}},\
  }\bibfield  {title} {\enquote {\bibinfo {title} {Accelerated methods for
  direct computation of fusion alpha particle losses within, stellarator
  optimization},}\ }\href@noop {} {\bibfield  {journal} {\bibinfo  {journal}
  {Journal of Plasma Physics}\ }\textbf {\bibinfo {volume} {86}},\ \bibinfo
  {pages} {815860201} (\bibinfo {year} {2020})}\BibitemShut {NoStop}%
\bibitem [{\citenamefont {Paul}\ \emph {et~al.}(2022)\citenamefont {Paul},
  \citenamefont {Bhattacharjee}, \citenamefont {Landreman}, \citenamefont
  {Alex}, \citenamefont {Velasco},\ and\ \citenamefont
  {Nies}}]{paul2022energetic}%
  \BibitemOpen
  \bibfield  {author} {\bibinfo {author} {\bibfnamefont {E.}~\bibnamefont
  {Paul}}, \bibinfo {author} {\bibfnamefont {A.}~\bibnamefont {Bhattacharjee}},
  \bibinfo {author} {\bibfnamefont {M.}~\bibnamefont {Landreman}}, \bibinfo
  {author} {\bibfnamefont {D.}~\bibnamefont {Alex}}, \bibinfo {author}
  {\bibfnamefont {J.}~\bibnamefont {Velasco}}, \ and\ \bibinfo {author}
  {\bibfnamefont {R.}~\bibnamefont {Nies}},\ }\bibfield  {title} {\enquote
  {\bibinfo {title} {Energetic particle loss mechanisms in reactor-scale
  equilibria close to quasisymmetry},}\ }\href@noop {} {\bibfield  {journal}
  {\bibinfo  {journal} {Nuclear Fusion}\ }\textbf {\bibinfo {volume} {62}},\
  \bibinfo {pages} {126054} (\bibinfo {year} {2022})}\BibitemShut {NoStop}%
\bibitem [{\citenamefont {Boozer}(1981)}]{boozer1981plasma}%
  \BibitemOpen
  \bibfield  {author} {\bibinfo {author} {\bibfnamefont {A.~H.}\ \bibnamefont
  {Boozer}},\ }\bibfield  {title} {\enquote {\bibinfo {title} {Plasma
  equilibrium with rational magnetic surfaces},}\ }\href@noop {} {\bibfield
  {journal} {\bibinfo  {journal} {The Physics of Fluids}\ }\textbf {\bibinfo
  {volume} {24}},\ \bibinfo {pages} {1999--2003} (\bibinfo {year}
  {1981})}\BibitemShut {NoStop}%
\bibitem [{\citenamefont {Xanthopoulos}\ \emph {et~al.}(2009)\citenamefont
  {Xanthopoulos}, \citenamefont {Cooper}, \citenamefont {Jenko}, \citenamefont
  {Turkin}, \citenamefont {Runov},\ and\ \citenamefont
  {Geiger}}]{xanthopoulos2009geometry}%
  \BibitemOpen
  \bibfield  {author} {\bibinfo {author} {\bibfnamefont {P.}~\bibnamefont
  {Xanthopoulos}}, \bibinfo {author} {\bibfnamefont {W.~A.}\ \bibnamefont
  {Cooper}}, \bibinfo {author} {\bibfnamefont {F.}~\bibnamefont {Jenko}},
  \bibinfo {author} {\bibfnamefont {Y.}~\bibnamefont {Turkin}}, \bibinfo
  {author} {\bibfnamefont {A.}~\bibnamefont {Runov}}, \ and\ \bibinfo {author}
  {\bibfnamefont {J.}~\bibnamefont {Geiger}},\ }\bibfield  {title} {\enquote
  {\bibinfo {title} {A geometry interface for gyrokinetic microturbulence
  investigations in toroidal configurations},}\ }\href@noop {} {\bibfield
  {journal} {\bibinfo  {journal} {Physics of Plasmas}\ }\textbf {\bibinfo
  {volume} {16}},\ \bibinfo {pages} {082303} (\bibinfo {year}
  {2009})}\BibitemShut {NoStop}%
\bibitem [{\citenamefont {Jenko}, \citenamefont {Dorland},\ and\ \citenamefont
  {Hammett}(2001)}]{jenko2001critical}%
  \BibitemOpen
  \bibfield  {author} {\bibinfo {author} {\bibfnamefont {F.}~\bibnamefont
  {Jenko}}, \bibinfo {author} {\bibfnamefont {W.}~\bibnamefont {Dorland}}, \
  and\ \bibinfo {author} {\bibfnamefont {G.}~\bibnamefont {Hammett}},\
  }\bibfield  {title} {\enquote {\bibinfo {title} {Critical gradient formula
  for toroidal electron temperature gradient modes},}\ }\href@noop {}
  {\bibfield  {journal} {\bibinfo  {journal} {Physics of Plasmas}\ }\textbf
  {\bibinfo {volume} {8}},\ \bibinfo {pages} {4096--4104} (\bibinfo {year}
  {2001})}\BibitemShut {NoStop}%
\bibitem [{\citenamefont {Landreman}\ \emph {et~al.}(2021)\citenamefont
  {Landreman}, \citenamefont {Medasani}, \citenamefont {Wechsung},
  \citenamefont {Giuliani}, \citenamefont {Jorge},\ and\ \citenamefont
  {Zhu}}]{landreman2021simsopt}%
  \BibitemOpen
  \bibfield  {author} {\bibinfo {author} {\bibfnamefont {M.}~\bibnamefont
  {Landreman}}, \bibinfo {author} {\bibfnamefont {B.}~\bibnamefont {Medasani}},
  \bibinfo {author} {\bibfnamefont {F.}~\bibnamefont {Wechsung}}, \bibinfo
  {author} {\bibfnamefont {A.}~\bibnamefont {Giuliani}}, \bibinfo {author}
  {\bibfnamefont {R.}~\bibnamefont {Jorge}}, \ and\ \bibinfo {author}
  {\bibfnamefont {C.}~\bibnamefont {Zhu}},\ }\bibfield  {title} {\enquote
  {\bibinfo {title} {Simsopt: A flexible framework for stellarator
  optimization},}\ }\href@noop {} {\bibfield  {journal} {\bibinfo  {journal}
  {Journal of Open Source Software}\ }\textbf {\bibinfo {volume} {6}},\
  \bibinfo {pages} {3525} (\bibinfo {year} {2021})}\BibitemShut {NoStop}%
\bibitem [{\citenamefont {Hegna}(2000)}]{hegna2000local}%
  \BibitemOpen
  \bibfield  {author} {\bibinfo {author} {\bibfnamefont {C.}~\bibnamefont
  {Hegna}},\ }\bibfield  {title} {\enquote {\bibinfo {title} {Local
  three-dimensional magnetostatic equilibria},}\ }\href@noop {} {\bibfield
  {journal} {\bibinfo  {journal} {Physics of Plasmas}\ }\textbf {\bibinfo
  {volume} {7}},\ \bibinfo {pages} {3921--3928} (\bibinfo {year}
  {2000})}\BibitemShut {NoStop}%
\bibitem [{\citenamefont {Gerard}\ \emph {et~al.}(2023)\citenamefont {Gerard},
  \citenamefont {Geiger}, \citenamefont {Pueschel}, \citenamefont {Bader},
  \citenamefont {Hegna}, \citenamefont {Faber}, \citenamefont {Terry},
  \citenamefont {Kumar},\ and\ \citenamefont {Schmitt}}]{Gerard_2023}%
  \BibitemOpen
  \bibfield  {author} {\bibinfo {author} {\bibfnamefont {M.~J.}\ \bibnamefont
  {Gerard}}, \bibinfo {author} {\bibfnamefont {B.}~\bibnamefont {Geiger}},
  \bibinfo {author} {\bibfnamefont {M.~J.}\ \bibnamefont {Pueschel}}, \bibinfo
  {author} {\bibfnamefont {A.}~\bibnamefont {Bader}}, \bibinfo {author}
  {\bibfnamefont {C.~C.}\ \bibnamefont {Hegna}}, \bibinfo {author}
  {\bibfnamefont {B.~J.}\ \bibnamefont {Faber}}, \bibinfo {author}
  {\bibfnamefont {P.~W.}\ \bibnamefont {Terry}}, \bibinfo {author}
  {\bibfnamefont {S.~T.~A.}\ \bibnamefont {Kumar}}, \ and\ \bibinfo {author}
  {\bibfnamefont {J.~C.}\ \bibnamefont {Schmitt}},\ }\bibfield  {title}
  {\enquote {\bibinfo {title} {Optimizing the hsx stellarator for
  microinstability by coil-current adjustments},}\ }\href {\doibase
  10.1088/1741-4326/acc1f6} {\bibfield  {journal} {\bibinfo  {journal}
  {Nucl.~Fusion}\ }\textbf {\bibinfo {volume} {63}},\ \bibinfo {pages} {056004}
  (\bibinfo {year} {2023})}\BibitemShut {NoStop}%
\bibitem [{\citenamefont {Calvetti}\ \emph {et~al.}(2000)\citenamefont
  {Calvetti}, \citenamefont {Golub}, \citenamefont {Gragg},\ and\ \citenamefont
  {Reichel}}]{calvetti2000computation}%
  \BibitemOpen
  \bibfield  {author} {\bibinfo {author} {\bibfnamefont {D.}~\bibnamefont
  {Calvetti}}, \bibinfo {author} {\bibfnamefont {G.}~\bibnamefont {Golub}},
  \bibinfo {author} {\bibfnamefont {W.}~\bibnamefont {Gragg}}, \ and\ \bibinfo
  {author} {\bibfnamefont {L.}~\bibnamefont {Reichel}},\ }\bibfield  {title}
  {\enquote {\bibinfo {title} {Computation of gauss-kronrod quadrature
  rules},}\ }\href@noop {} {\bibfield  {journal} {\bibinfo  {journal}
  {Mathematics of computation}\ }\textbf {\bibinfo {volume} {69}},\ \bibinfo
  {pages} {1035--1052} (\bibinfo {year} {2000})}\BibitemShut {NoStop}%
\bibitem [{\citenamefont {Gander}\ and\ \citenamefont
  {Gautschi}(2000)}]{gander2000adaptive}%
  \BibitemOpen
  \bibfield  {author} {\bibinfo {author} {\bibfnamefont {W.}~\bibnamefont
  {Gander}}\ and\ \bibinfo {author} {\bibfnamefont {W.}~\bibnamefont
  {Gautschi}},\ }\bibfield  {title} {\enquote {\bibinfo {title} {Adaptive
  quadrature—revisited},}\ }\href@noop {} {\bibfield  {journal} {\bibinfo
  {journal} {BIT Numerical Mathematics}\ }\textbf {\bibinfo {volume} {40}},\
  \bibinfo {pages} {84--101} (\bibinfo {year} {2000})}\BibitemShut {NoStop}%
\bibitem [{\citenamefont {Gonnet}(2012)}]{gonnet2012review}%
  \BibitemOpen
  \bibfield  {author} {\bibinfo {author} {\bibfnamefont {P.}~\bibnamefont
  {Gonnet}},\ }\bibfield  {title} {\enquote {\bibinfo {title} {A review of
  error estimation in adaptive quadrature},}\ }\href@noop {} {\bibfield
  {journal} {\bibinfo  {journal} {ACM Computing Surveys (CSUR)}\ }\textbf
  {\bibinfo {volume} {44}},\ \bibinfo {pages} {1--36} (\bibinfo {year}
  {2012})}\BibitemShut {NoStop}%
\bibitem [{\citenamefont {Velasco}\ \emph {et~al.}(2020)\citenamefont
  {Velasco}, \citenamefont {Calvo}, \citenamefont {Parra},\ and\ \citenamefont
  {Garc{\'\i}a-Rega{\~n}a}}]{velasco2020knosos}%
  \BibitemOpen
  \bibfield  {author} {\bibinfo {author} {\bibfnamefont {J.~L.}\ \bibnamefont
  {Velasco}}, \bibinfo {author} {\bibfnamefont {I.}~\bibnamefont {Calvo}},
  \bibinfo {author} {\bibfnamefont {F.~I.}\ \bibnamefont {Parra}}, \ and\
  \bibinfo {author} {\bibfnamefont {J.}~\bibnamefont
  {Garc{\'\i}a-Rega{\~n}a}},\ }\bibfield  {title} {\enquote {\bibinfo {title}
  {Knosos: a fast orbit-averaging neoclassical code for stellarator
  geometry},}\ }\href@noop {} {\bibfield  {journal} {\bibinfo  {journal}
  {Journal of Computational Physics}\ }\textbf {\bibinfo {volume} {418}},\
  \bibinfo {pages} {109512} (\bibinfo {year} {2020})}\BibitemShut {NoStop}%
\bibitem [{\citenamefont {Corduneanu}(2009)}]{corduneanu2009almost}%
  \BibitemOpen
  \bibfield  {author} {\bibinfo {author} {\bibfnamefont {C.}~\bibnamefont
  {Corduneanu}},\ }\href@noop {} {\emph {\bibinfo {title} {Almost periodic
  oscillations and waves}}}\ (\bibinfo  {publisher} {Springer Science \&
  Business Media},\ \bibinfo {year} {2009})\BibitemShut {NoStop}%
\bibitem [{\citenamefont {Oka}(2022)}]{oka2022space}%
  \BibitemOpen
  \bibfield  {author} {\bibinfo {author} {\bibfnamefont {T.}~\bibnamefont
  {Oka}},\ }\bibfield  {title} {\enquote {\bibinfo {title} {Space--time
  arithmetic quasi-periodic homogenization for damped wave equations},}\
  }\href@noop {} {\bibfield  {journal} {\bibinfo  {journal} {Results in Applied
  Mathematics}\ }\textbf {\bibinfo {volume} {15}},\ \bibinfo {pages} {100310}
  (\bibinfo {year} {2022})}\BibitemShut {NoStop}%
\bibitem [{\citenamefont {Rosenbluth}(1968)}]{rosenbluth1968}%
  \BibitemOpen
  \bibfield  {author} {\bibinfo {author} {\bibfnamefont {M.~N.}\ \bibnamefont
  {Rosenbluth}},\ }\bibfield  {title} {\enquote {\bibinfo {title}
  {Low‐frequency limit of interchange instability},}\ }\href {\doibase
  10.1063/1.1692009} {\bibfield  {journal} {\bibinfo  {journal} {The Physics of
  Fluids}\ }\textbf {\bibinfo {volume} {11}},\ \bibinfo {pages} {869--872}
  (\bibinfo {year} {1968})},\ \Eprint
  {http://arxiv.org/abs/https://aip.scitation.org/doi/pdf/10.1063/1.1692009}
  {https://aip.scitation.org/doi/pdf/10.1063/1.1692009} \BibitemShut {NoStop}%
\bibitem [{\citenamefont {Fritsch}\ and\ \citenamefont
  {Butland}(1984)}]{fritsch1984method}%
  \BibitemOpen
  \bibfield  {author} {\bibinfo {author} {\bibfnamefont {F.~N.}\ \bibnamefont
  {Fritsch}}\ and\ \bibinfo {author} {\bibfnamefont {J.}~\bibnamefont
  {Butland}},\ }\bibfield  {title} {\enquote {\bibinfo {title} {A method for
  constructing local monotone piecewise cubic interpolants},}\ }\href@noop {}
  {\bibfield  {journal} {\bibinfo  {journal} {SIAM journal on scientific and
  statistical computing}\ }\textbf {\bibinfo {volume} {5}},\ \bibinfo {pages}
  {300--304} (\bibinfo {year} {1984})}\BibitemShut {NoStop}%
\bibitem [{\citenamefont {Takahasi}\ and\ \citenamefont
  {Mori}(1974)}]{takahasi1974double}%
  \BibitemOpen
  \bibfield  {author} {\bibinfo {author} {\bibfnamefont {H.}~\bibnamefont
  {Takahasi}}\ and\ \bibinfo {author} {\bibfnamefont {M.}~\bibnamefont
  {Mori}},\ }\bibfield  {title} {\enquote {\bibinfo {title} {Double exponential
  formulas for numerical integration},}\ }\href@noop {} {\bibfield  {journal}
  {\bibinfo  {journal} {Publications of the Research Institute for Mathematical
  Sciences}\ }\textbf {\bibinfo {volume} {9}},\ \bibinfo {pages} {721--741}
  (\bibinfo {year} {1974})}\BibitemShut {NoStop}%
\bibitem [{\citenamefont {Connor}, \citenamefont {Hastie},\ and\ \citenamefont
  {Martin}(1983)}]{connor1983effect}%
  \BibitemOpen
  \bibfield  {author} {\bibinfo {author} {\bibfnamefont {J.}~\bibnamefont
  {Connor}}, \bibinfo {author} {\bibfnamefont {R.}~\bibnamefont {Hastie}}, \
  and\ \bibinfo {author} {\bibfnamefont {T.}~\bibnamefont {Martin}},\
  }\bibfield  {title} {\enquote {\bibinfo {title} {Effect of pressure gradients
  on the bounce-averaged particle drifts in a tokamak},}\ }\href@noop {}
  {\bibfield  {journal} {\bibinfo  {journal} {Nuclear fusion}\ }\textbf
  {\bibinfo {volume} {23}},\ \bibinfo {pages} {1702} (\bibinfo {year}
  {1983})}\BibitemShut {NoStop}%
\bibitem [{\citenamefont {Duff}()}]{JDuff_NE3DLE}%
  \BibitemOpen
  \bibfield  {author} {\bibinfo {author} {\bibfnamefont {J.}~\bibnamefont
  {Duff}},\ }\href@noop {} {\enquote {\bibinfo {title} {{NE3DLE} code},}\
  }\bibinfo {howpublished} {\url{https://gitlab.com/jduff2/NE3DLE}},\ \bibinfo
  {note} {accessed: 2023-02-17}\BibitemShut {NoStop}%
\bibitem [{\citenamefont {Zarnstorff}\ \emph {et~al.}(2001)\citenamefont
  {Zarnstorff}, \citenamefont {Berry}, \citenamefont {Brooks}, \citenamefont
  {Fredrickson}, \citenamefont {Fu}, \citenamefont {Hirshman}, \citenamefont
  {Hudson}, \citenamefont {Ku}, \citenamefont {Lazarus}, \citenamefont
  {Mikkelsen} \emph {et~al.}}]{zarnstorff2001physics}%
  \BibitemOpen
  \bibfield  {author} {\bibinfo {author} {\bibfnamefont {M.}~\bibnamefont
  {Zarnstorff}}, \bibinfo {author} {\bibfnamefont {L.}~\bibnamefont {Berry}},
  \bibinfo {author} {\bibfnamefont {A.}~\bibnamefont {Brooks}}, \bibinfo
  {author} {\bibfnamefont {E.}~\bibnamefont {Fredrickson}}, \bibinfo {author}
  {\bibfnamefont {G.}~\bibnamefont {Fu}}, \bibinfo {author} {\bibfnamefont
  {S.}~\bibnamefont {Hirshman}}, \bibinfo {author} {\bibfnamefont
  {S.}~\bibnamefont {Hudson}}, \bibinfo {author} {\bibfnamefont
  {L.}~\bibnamefont {Ku}}, \bibinfo {author} {\bibfnamefont {E.}~\bibnamefont
  {Lazarus}}, \bibinfo {author} {\bibfnamefont {D.}~\bibnamefont {Mikkelsen}},
  \emph {et~al.},\ }\bibfield  {title} {\enquote {\bibinfo {title} {Physics of
  the compact advanced stellarator ncsx},}\ }\href@noop {} {\bibfield
  {journal} {\bibinfo  {journal} {Plasma Physics and Controlled Fusion}\
  }\textbf {\bibinfo {volume} {43}},\ \bibinfo {pages} {A237} (\bibinfo {year}
  {2001})}\BibitemShut {NoStop}%
\bibitem [{\citenamefont {Antonsen~Jr}\ and\ \citenamefont
  {Lane}(1980)}]{antonsen1980kinetic}%
  \BibitemOpen
  \bibfield  {author} {\bibinfo {author} {\bibfnamefont {T.~M.}\ \bibnamefont
  {Antonsen~Jr}}\ and\ \bibinfo {author} {\bibfnamefont {B.}~\bibnamefont
  {Lane}},\ }\bibfield  {title} {\enquote {\bibinfo {title} {Kinetic equations
  for low frequency instabilities in inhomogeneous plasmas},}\ }\href@noop {}
  {\bibfield  {journal} {\bibinfo  {journal} {The Physics of Fluids}\ }\textbf
  {\bibinfo {volume} {23}},\ \bibinfo {pages} {1205--1214} (\bibinfo {year}
  {1980})}\BibitemShut {NoStop}%
\bibitem [{\citenamefont {Catto}, \citenamefont {Tang},\ and\ \citenamefont
  {Baldwin}(1981)}]{catto1981generalized}%
  \BibitemOpen
  \bibfield  {author} {\bibinfo {author} {\bibfnamefont {P.}~\bibnamefont
  {Catto}}, \bibinfo {author} {\bibfnamefont {W.}~\bibnamefont {Tang}}, \ and\
  \bibinfo {author} {\bibfnamefont {D.}~\bibnamefont {Baldwin}},\ }\bibfield
  {title} {\enquote {\bibinfo {title} {Generalized gyrokinetics},}\ }\href@noop
  {} {\bibfield  {journal} {\bibinfo  {journal} {Plasma Physics}\ }\textbf
  {\bibinfo {volume} {23}},\ \bibinfo {pages} {639} (\bibinfo {year}
  {1981})}\BibitemShut {NoStop}%
\bibitem [{\citenamefont {Villard}\ \emph {et~al.}(2002)\citenamefont
  {Villard}, \citenamefont {Bottino}, \citenamefont {Sauter},\ and\
  \citenamefont {Vaclavik}}]{villard2002radial}%
  \BibitemOpen
  \bibfield  {author} {\bibinfo {author} {\bibfnamefont {L.}~\bibnamefont
  {Villard}}, \bibinfo {author} {\bibfnamefont {A.}~\bibnamefont {Bottino}},
  \bibinfo {author} {\bibfnamefont {O.}~\bibnamefont {Sauter}}, \ and\ \bibinfo
  {author} {\bibfnamefont {J.}~\bibnamefont {Vaclavik}},\ }\bibfield  {title}
  {\enquote {\bibinfo {title} {Radial electric fields and global electrostatic
  microinstabilities in tokamaks and stellarators},}\ }\href@noop {} {\bibfield
   {journal} {\bibinfo  {journal} {Physics of Plasmas}\ }\textbf {\bibinfo
  {volume} {9}},\ \bibinfo {pages} {2684--2691} (\bibinfo {year}
  {2002})}\BibitemShut {NoStop}%
\bibitem [{\citenamefont {Xanthopoulos}\ \emph {et~al.}(2016)\citenamefont
  {Xanthopoulos}, \citenamefont {Plunk}, \citenamefont {Zocco},\ and\
  \citenamefont {Helander}}]{xanthopoulos2016}%
  \BibitemOpen
  \bibfield  {author} {\bibinfo {author} {\bibfnamefont {P.}~\bibnamefont
  {Xanthopoulos}}, \bibinfo {author} {\bibfnamefont {G.~G.}\ \bibnamefont
  {Plunk}}, \bibinfo {author} {\bibfnamefont {A.}~\bibnamefont {Zocco}}, \ and\
  \bibinfo {author} {\bibfnamefont {P.}~\bibnamefont {Helander}},\ }\bibfield
  {title} {\enquote {\bibinfo {title} {Intrinsic turbulence stabilization in a
  stellarator},}\ }\href {\doibase 10.1103/PhysRevX.6.021033} {\bibfield
  {journal} {\bibinfo  {journal} {Phys. Rev. X}\ }\textbf {\bibinfo {volume}
  {6}},\ \bibinfo {pages} {021033} (\bibinfo {year} {2016})}\BibitemShut
  {NoStop}%
\bibitem [{Note1()}]{Note1}%
  \BibitemOpen
  \bibinfo {note} {An important corollary is that the stability criterion is
  strictly only true in exactly omnigenous systems.}\BibitemShut {Stop}%
\bibitem [{\citenamefont {Mackenbach}, \citenamefont {Duff},\ and\
  \citenamefont {Gerard}()}]{BAD_code}%
  \BibitemOpen
  \bibfield  {author} {\bibinfo {author} {\bibfnamefont {R.}~\bibnamefont
  {Mackenbach}}, \bibinfo {author} {\bibfnamefont {J.}~\bibnamefont {Duff}}, \
  and\ \bibinfo {author} {\bibfnamefont {M.}~\bibnamefont {Gerard}},\
  }\href@noop {} {\enquote {\bibinfo {title} {{BAD} code},}\ }\bibinfo
  {howpublished} {\url{https://github.com/RalfMackenbach/BAD}},\ \bibinfo
  {note} {accessed: 2023-04-11}\BibitemShut {NoStop}%
\end{thebibliography}%
\end{document}